\documentclass[10pt,journal,compsoc]{IEEEtran}
\ifCLASSOPTIONcompsoc
  \usepackage[nocompress]{cite}
\else
  \usepackage{cite}
\fi
\usepackage[dvipsnames]{xcolor}
\usepackage[export]{adjustbox}
\usepackage[T1]{fontenc}
\usepackage{textcomp}
\usepackage{enumitem}
\usepackage{import}
\usepackage{multirow}
\usepackage{graphicx}
\usepackage{booktabs}
\usepackage{changepage}
\usepackage[export]{adjustbox}
\usepackage[
  separate-uncertainty = true,
  multi-part-units = repeat
]{siunitx}
\usepackage[numbers]{natbib}
\usepackage{booktabs}
\usepackage{graphicx}
\usepackage{soul}
\usepackage[framemethod=TikZ]{mdframed}
\usepackage{filecontents}
\usepackage{caption}
\usepackage{subcaption}
\usepackage{float}
\usepackage{vwcol}  
\restylefloat{table}
\usepackage{adjustbox}

\usepackage{rotating}
\usepackage{longtable}
\usepackage{multirow}
\usepackage{lscape} 
\usepackage{rotating}
\usepackage[most]{tcolorbox}
\usepackage[table,xcdraw]{xcolor}
\usepackage{algorithmic}
\usepackage[hidelinks]{hyperref}

\definecolor{custom-gray}{cmyk}{0, 0, 0, 0.7, 1.00}
\newtcbtheorem[no counter]{Summary}{\hskip-0.97em}{enhanced,drop shadow={black!50!white},
  coltitle=white,
  top=0.15in,
  attach boxed title to top left=
  {xshift=1.5em,yshift=-\tcboxedtitleheight/2},
  boxed title style={size=small,colback=custom-gray}
}{summary}

\ifCLASSINFOpdf
\else
\fi
\usepackage{eso-pic}
\AddToShipoutPictureFG*{\AtPageLowerLeft{\put(\LenToUnit{0.08\paperwidth},\LenToUnit{0.022\paperheight}){\parbox{0.84\paperwidth}{\scriptsize\centering \copyright~2026 IEEE. Personal use of this material is permitted. Permission from IEEE must be obtained for all other uses, in any current or future media, including reprinting/republishing this material for advertising or promotional purposes, creating new collective works, for resale or redistribution to servers or lists, or reuse of any copyrighted component of this work in other works.}}}}
\begin{document}
%
\title{Retrieval Supersedes Scale: Efficient Log Level Prediction via Multiplex Socio-Technical Context}
%
%
%
%
\newcommand{\RQI}{How does the integration of ownership and functionality signals via multiplex clustering impact the predictive performance of LLM-powered LLPs?}
\newcommand{\RQII}{What is the individual contribution of functionality and ownership signals?}
\newcommand{\RQIII}{Can multiplex Retrieval reduce the computational cost of LLM-powered log level prediction?}
\newcommand{\RQIV}{Can multiplex Retrieval reduce the computational cost of LLM-powered logging level prediction?}

\author{Youssef Esseddiq~Ouatiti,
        Mohammed~Sayagh,
        Bram Adams,
        and~Ahmed~E.~Hassan
\thanks{Y.E. Ouatiti, B. Adams and A.E. Hassan are with the School of Computing, Queen's University, Kingston, ON, Canada.}
\thanks{M. Sayagh is with the \'Ecole de Technologie Sup\'erieure, Universit\'e de Qu\'ebec, Montreal, QC, Canada.}
        
}


%
%


\IEEEtitleabstractindextext{%
\begin{abstract}
Developers insert logging statements in the source code to log its execution. The generated logs are essential for developers in predicting, detecting and troubleshooting system anomalies. A key part of these statements is the log level (e.g., \texttt{DEBUG}, \texttt{INFO}, \texttt{WARN}, \texttt{ERROR}, \texttt{FATAL}), which determines whether a log message is recorded. Choosing an appropriate log level is a complex trade-off as levels set too low (\texttt{TRACE} or \texttt{DEBUG}) mask important information in production, while levels set too high (\texttt{ERROR} or \texttt{FATAL}) create noisy logs and impose extra I/O and storage overhead. A large body of research investigates how to predict the accurate log level for a logging statement leveraging AI approaches. For instance, recent work leverages Large Language Models (LLMs) via In-Context Learning (ICL) to infer the suitable log level for a logging statement. However, current approaches typically augment the prediction prompt with randomly sampled examples. We argue that this diversity-based approach overlooks a critical reality pertaining to the socio-technical nature of logging, where conventions may vary significantly by component and development teams. Consequently, off-the-shelf LLMs using random examples are reported to achieve median AUCs of only 0.56 to 0.76 (across evaluated projects), leaving room for improvement without resorting to fine-tuning techniques. In this paper, we propose a retrieval approach for LLM-powered log level prediction that prioritizes context relevance over random selection of ICL examples and basic similarity approaches typically used in ICL retrieval. Our ICL RAG-based method retrieves logging examples relevant along two socio-technical dimensions: (i) functionality (code performing similar operations) and (ii) ownership (code maintained by the same set of developers). Evaluating four different LLMs across ten popular Java projects, we find that this retrieval improves AUC over random baselines by 0.07 (median). Combining both signals yields median AUCs of 0.83 to 0.97. Furthermore, our ablation studies reveal that each signal improves on random retrieval alone, and that combining them outperforms the stronger of the two on all ten projects. Our AUC results also significantly exceed state-of-the-art standard supervised predictors (e.g., DeepLV) by a median between 0.12 and 0.19 across our evaluated projects. Our findings highlight the importance of leveraging socio-technical signals such as code functionality and ownership to inform LLM-powered log level prediction.  
\end{abstract}
}

\maketitle

\section{Introduction}
Developers insert logging statements into the source code of software systems to capture essential runtime information, facilitating debugging, system observability, and anomaly prediction, detection, and diagnosis~\cite{Li:2020,Shang:2015,Yuan:2012}. A typical logging statement includes a log level (e.g., \texttt{DEBUG}), and a textual message with or without variables. Log levels are typically ordered by verbosity from \texttt{DEBUG} (most verbose) to \texttt{FATAL} (least verbose) in Java logging libraries. Such log levels control the volume of logged data, as during software operation, only log messages at or above a set threshold are recorded, while more verbose messages are excluded. Operators set this threshold through configuration, independently of the levels written in the code.

However, accurately determining the appropriate log level for each logging statement remains a challenging activity for developers~\cite{Li:2017,Pecchia:2015}, who typically make initial poor log level choices~\cite{Li:2020,Oliner:2012}. Suboptimal log level decisions have significant practical consequences as a too low log level drastically reduces software observability, complicates debugging, and prolongs failure diagnosis~\cite{Yuan:2012}, while an excessive log level introduces performance overhead, inflates storage, and overwhelms developers with redundant information~\cite{Yuan:2014}.

To assist developers in log level decision-making process, researchers have proposed various automated Log Level Predictors (LLPs)~\cite{Anu:2019,Heng:2024,Kim:2019,Li:2017,Liu:2022,LiZ:2021a}. Early approaches~\cite{Anu:2019,Li:2017,Kim:2019} relied on engineered software features (e.g., cyclomatic complexity) to train shallow classifiers such as ordinal logistic regression. Subsequent research shifted toward deep learning, leveraging learned embeddings of logging statements and their surrounding code context to capture functionality nuances~\cite{Liu:2022,LiZ:2021a}. Most recently, the advent of generative AI has spurred the adoption of Large Language Models (LLMs) for log level prediction, utilizing paradigms ranging from zero-shot inference to few-shot In-Context-Learning (ICL) and fine-tuning (e.g., LoRA)~\cite{Heng:2024}.

Current LLM-powered log level prediction approaches typically rely on random or stratified sampling for in-context examples, a practice that implicitly favors data diversity over contextual relevance~\cite{Heng:2024,Ma:2024}. While often chosen for simplicity, this design is supported by the assumed benefits of data diversity in deep learning~\cite{Gong:2019}, including (1) improved robustness on unseen inputs, (2) mitigation of selection bias, and (3) coverage of all output classes within the limited context window. Consequently, diverse, randomized sampling is widely treated as a reasonable default for conditioning frozen LLMs. 

We argue that this approach (i.e., diversity-first) overlooks key aspects that are fundamental to the logging practice~\cite{Ouatiti:2023}. Specifically, projects are not a uniform entity, with a single, consistent logging strategy. In reality, distinct components or subsystems often exhibit different logging practices influenced by factors such as code functionality (i.e., what the code does) and developer ownership (i.e., who maintains or frequently changes a file). We hypothesize in this paper that leveraging these internal distinctions can allow existing LLM-powered LLPs to use the most contextually relevant software knowledge, thereby enhancing their predictive accuracy and practical utility.

For example, the SIG-Auth team\footnote{\url{https://github.com/kubernetes/community/tree/master/sig-auth}} in the Kubernetes community maintains authentication logic dispersed across the API Server and the Kubelet. These files sit in different components and have distinct functionality, yet they may share logging conventions because the same people write them. Similarly, Hadoop's \textit{RetryInvocationHandler} and Yarn's \textit{RequestHedgingRMFailoverProxyProvider} implement the same retry duty in different components, and both log each attempt at \texttt{DEBUG} before escalating once the retries are exhausted, differing only in the terminal severity that each component's failure warrants. Logging conventions are therefore shaped by both shared ownership and functional similarity, and omitting either signal may limit the predictive accuracy of LLM-powered LLPs.
We select functionality and ownership as a complementary pair rather than as an exhaustive model of socio-technical context. Functionality represents the technical responsibilities and observability needs of the code, while ownership represents the social and evolutionary context in which local logging conventions develop. Both signals can be derived consistently at file granularity across the studied projects and represented as aligned layers of the same multiplex graph. 

In this paper, we investigate the impact of two socio-technical signals (i.e., code functionality and ownership) on the performance of in-context-learning LLM-powered log level prediction. We propose a methodology that extracts and fuses two distinct socio-technical signals (i.e., code functionality and ownership) into the retrieval process. Specifically, we model a studied software project as a two-layer multiplex graph~\cite{Battiston:2014,Domenico:2013}, where nodes represent source files and layers represent the distinct relationship types. While the first layer encodes functionality, where edges connect files with high functional similarity, the second layer encodes ownership, where edges connect files maintained by the same group of contributors. By applying community detection on this unified multiplex structure, we partition the project into coherent socio-technical clusters. Finally, for any given target file, we restrict the retrieval of in-context-learning examples exclusively to its assigned cluster, ensuring that the LLM is augmented with examples that are both functionally relevant and aligned with the team's logging conventions.

It is important to distinguish log level prediction from the broader task of end-to-end log generation. While generative tools aim to solve the ```cold start''' problem by synthesizing messages for unlogged code, they introduce significant risk when applied to existing logging statements as regenerating a full statement can alter developer-tuned message content or variable selection, degrading log quality and increasing code review burden~\cite{Rawte:2023}. In practice, developers often simply need to 'up-shift' or 'down-shift' the visibility of an existing log message without rewriting it. Therefore, log level prediction addresses a distinct calibration problem pertaining to the optimization of the trade-off between observability and overhead for developer-written log messages. Furthermore, our study concerns log level prediction with frozen, out-of-the-box LLMs, where the model parameters remain unchanged and only the in-context-learning examples vary. Although fine-tuning can adapt an LLM to a specific project logging data, it requires a significant amount of labeled training data and parameter optimization, which add training and maintenance costs. Our approach instead keeps the LLM frozen and adapts its context at inference time through carefully retrieved examples. This provides a lighter-weight form of context adaptation and allows us to isolate the contribution of retrieval under identical model and inference configuration.

To evaluate our approach, we answer the following RQs:

\textbf{RQ1. \RQI} Our approach, leveraging both functionality and ownership signals simultaneously via multiplex clustering, achieves a median AUC between 0.83 and 0.97 across our evaluated projects. The approach statistically significantly (Wilcoxon test, $\alpha = 0.01$) outperforms state-of-the-art approaches in LLM-powered log level prediction (in terms of AUC) by a median of 0.07. Moreover, this integration of socio-technical signals enables frozen LLMs to exceed the AUC of prior specialized supervised models (e.g., DeepLV: +0.12, Ordinal Regression: +0.19) while avoiding task-specific model training and ongoing retraining costs.

\textbf{RQ2. \RQII}
Leveraging in-context-learning examples retrieved from our file ownership clusters alone statistically significantly enhances the AUC of LLM-powered log level prediction by a median of 0.02 (Wilcoxon test, $\alpha=0.01$), compared to the state-of-the-art approaches in LLM-powered log level prediction. Furthermore, leveraging in-context-learning examples retrieved from our functionality-only clusters statistically significantly enhances the AUC of LLM-powered log level prediction by a median of 0.05 (Wilcoxon test, $\alpha = 0.01$), compared to the state-of-the-art approaches in LLM-powered log level prediction.

\textbf{RQ3. \RQIII} Our findings indicate that precise context retrieval is more critical than raw model scale. A multiplex-enhanced 7B model outperforms state-of-the-art commercial models, achieving an average AUC of 0.89, surpassing GPT-4o by 0.10 AUC and DeepSeek-V3.2 by 0.16 AUC. We demonstrate that scaling to larger local models (34B) yields diminishing returns, offering only marginal performance gains (+0.01 AUC) at a disproportionate operational cost increase (+322\%).

Our findings suggest the adoption of small, self-hosted and context-aware models over general-purpose commercial APIs for LLM-powered log level prediction. We observe that integrating socio-technical signals (i.e., ownership and functionality) enables smaller models (e.g., 7B) to outperform much larger baselines, proving that precise retrieval supersedes raw parameter scale. We suggest practitioners prioritize self-hosted, batched inference pipelines to balance accuracy with operational efficiency for log level prediction tasks.


\section{Background \& Related work}

\subsection{Log level prediction}
One direction in software logging research aims to automate the log level choice~\cite{Anu:2019,Heng:2024,Li:2017,LiZ:2021a}, as developers and empirical studies reported challenges with the log level choice activity. For instance, Li et al.~\cite{Li:2017} leveraged an ordinal logistic regression model, specifically designed to predict log levels using a set of static, change-related and historical code features (e.g., log message length, logging statement churn, number of revisions in history). Anu et al.~\cite{Anu:2019} introduced VerbosityLevelDirector, an automated approach based on random-forest designed to help developers assign appropriate log levels based on logging intention embedded within the code context. More recently, Li et al.~\cite{LiZ:2021a} introduced a deep learning take on log level prediction. Their approach leveraged a neural network architecture that encoded syntactic and semantic contexts surrounding logging statements as well as the contents of the logging statements, to predict an ordinal output corresponding to the different log levels. Heng et al.~\cite{Heng:2024} explored log level prediction using an LLM-powered approach, employing various LLMs with different sizes (e.g., CodeLlama-7B). Given a source-code snippet (where the logging statement is masked) as a context, and the content of the log message, the LLM predicts the appropriate log level by generating an appropriate textual response. 

That said, state-of-the-art efforts leveraging LLMs for log level prediction leverage random retrieval due to the ease of the approach and as a way to promote diversity in the examples provided to LLMs at inference time. In this paper, we investigate log level prediction approaches that take into consideration the socio-technical aspects of software systems (i.e., the code functionality and its ownership).

\subsection{In-context-learning for logging tasks}
Prior studies have leveraged in-context-learning~\cite{Brown:2020,Wang:2020} for logging automation to bridge the gap in domain knowledge by incorporating task-specific examples directly into the input prompt without parameter updates~\cite{Heng:2024,Li:2024,UniLog:2024}. However, the effectiveness of in-context-learning is often constrained by the limited context window, which restricts the number of demonstrations that can be utilized during inference. Such in-context-learning approach was evaluated for log level prediction~\cite{Heng:2024,Li:2024,UniLog:2024}. For instance, Xu et al.~\cite{UniLog:2024} introduced UniLog, an LLM-powered solution that employs in-context-learning to automatically generate log messages, recommend appropriate logging statement placements, and predict their verbosity levels. Heng et al.~\cite{Heng:2024} investigated log level prediction using various LLMs with different sizes and model architectures, and showed that even randomly picked in-context-learning examples can lift the performance of LLM-powered LLPs. Li et al.~\cite{Li:2024} built LogBench, a logging benchmark of code instances collected from GitHub repositories together with a transformed, unseen counterpart, and used it to evaluate 13 LLMs ranging from 60 million to 405 billion parameters on logging tasks, including log level prediction.

Our work is in line with this research direction as we evaluate in-context-learning approaches for log level prediction. Yet, instead of random in-context-learning retrieval, our study investigates how the selection of in-context-learning examples can be improved when taking into consideration the ownership and functionality relationships present within modern software systems. 
\section{Methodology}
In this section, we introduce an in-context-learning retrieval approach designed for log level prediction. Our approach provides relevant logging examples to LLMs to generate log levels for newly added or modified logging statements within source code, as shown in Figure~\ref{fig:methodology_overview}. The input of our approach is the file name and the specific logging message for which a log level prediction is to be predicted. The output is the predicted log level. To perform the prediction, we first select in-context logging examples from the cluster that is relevant both in terms of functionality and ownership to the logging statement in hand. We discuss the creation of these clusters in the following sections:  

\begin{figure}[t]
  \centering
  \includegraphics[width=\textwidth,keepaspectratio,height=6.0cm]{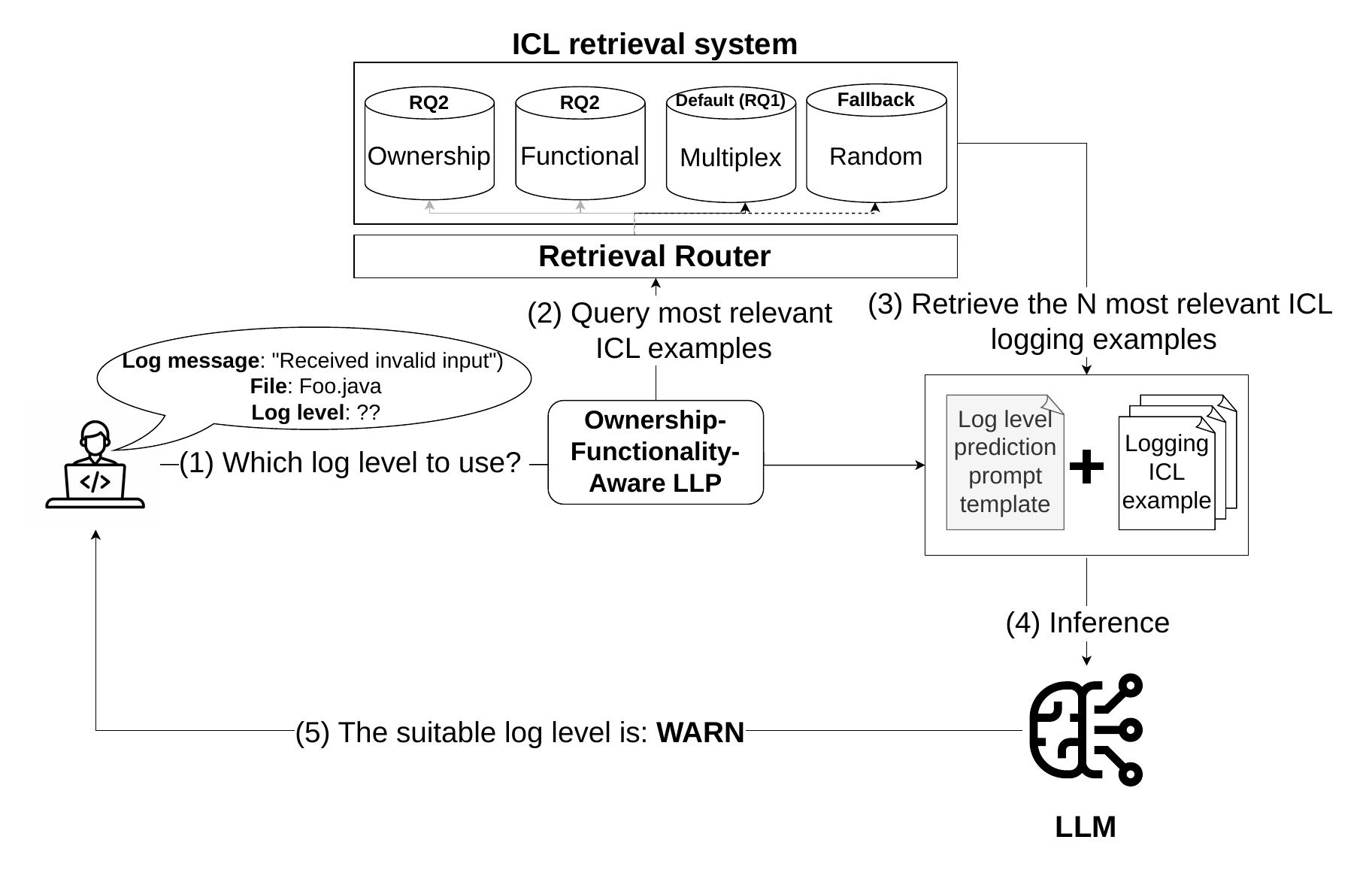}
  \caption{Overview of our Ownership-Functionality Aware retrieval approach for log level prediction.}
  \label{fig:methodology_overview}
\end{figure}

\subsection{Functionality-based clustering}
\label{sec:sem}
We leverage functionality-based clustering to group files based on their functional similarity, to identify functionally cohesive file clusters that are expected to share similar logging behaviors.

\begin{figure}[t]
\centering
\includegraphics[width=\textwidth,keepaspectratio,height=3.2cm]{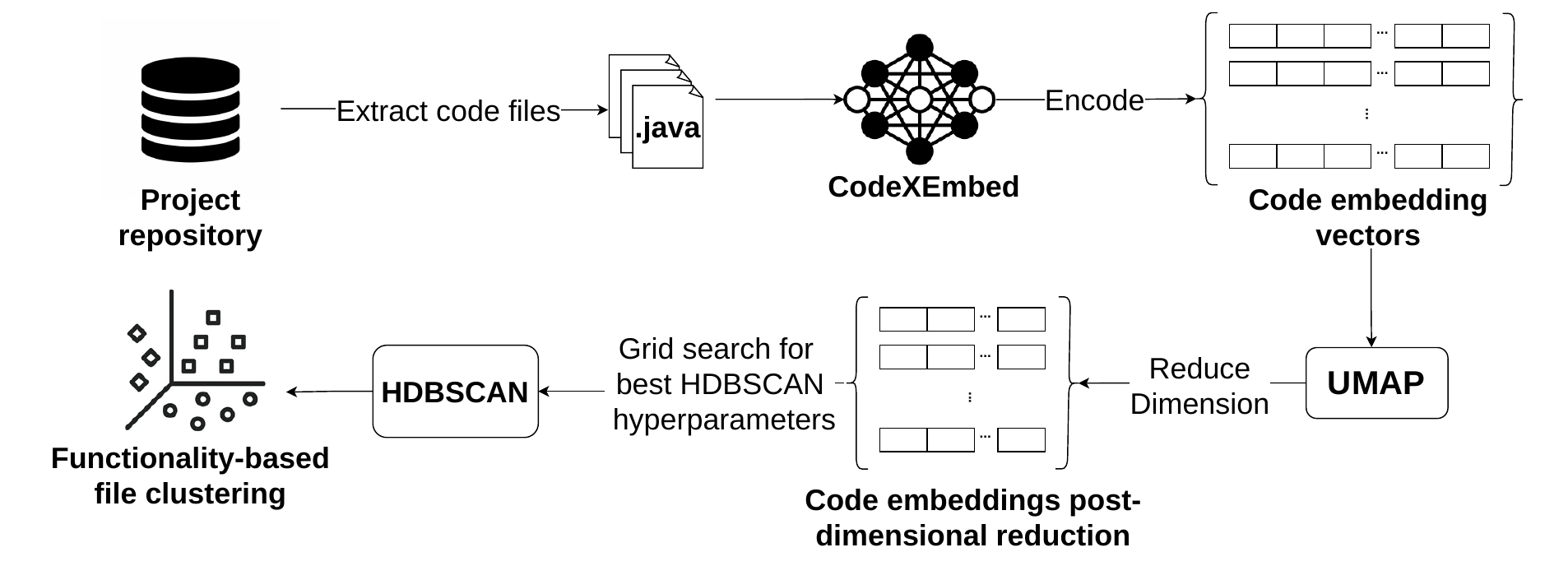}
\caption{Overview of the functionality-based clustering}
\label{fig:semantic_clustering}
\end{figure}

\textbf{Raw data construction}: To build our functionality-based clusters we follow the approach summarized in Figure~\ref{fig:semantic_clustering}. We first encode the source files of each project with a transformer-based embedding model, which produces one dense vector per file that reflects both the syntax and the functionality of its code~\cite{Chochlov:2022,Feng:2020,Liu:2024}. We chose the encoder-only transformer CodeXEmbed~\cite{Liu:2024} because it provides state-of-the-art performance on code-related embedding tasks, supports input sequences of up to 32K tokens --thus effectively capturing the full context of most source files in our studied projects-- and achieves these capabilities despite its modest size (2.6B parameters), making it computationally efficient and practical. After the encoding, each source file is represented as a 2,304-dimensional embedding vector capturing the file's functional information. Files exceeding the 32K token window (less than 5\% across projects) are partitioned at top-level method boundaries and encoded per method, then mean-pooled into a single file-level vector~\cite{Gunther:2024}, so that every file receives exactly one embedding and remains eligible for clustering regardless of its size.



\textbf{Dimensionality reduction}: As direct clustering on high-dimensional embeddings is computationally expensive and sensitive to noise, we first apply UMAP (Uniform Manifold Approximation and Projection)~\cite{Mcinnes:2020} to project the 2,304-dimensional file embeddings into a 50-dimensional space, similarly to prior work~\cite{Baligodugula:2025}. UMAP preserves the local neighborhood structure while substantially reducing the computational cost of the clustering.

\textbf{Functional-similarity clustering}: We then apply the density-based clustering algorithm HDBSCAN~\cite{Malzer:2020} to the reduced embeddings. Density-based clustering is a deliberate design choice for our setting. Unlike partitional methods, which force every file into a cluster, HDBSCAN discovers clusters of varying densities and explicitly labels low-density files as noise. This matters because source files in large projects are heterogeneous, and assigning a file to a functionally incoherent cluster would yield misleading in-context-learning examples. For each project, we select the two main HDBSCAN hyperparameters: (i) the minimum cluster size and (ii) the minimum number of samples, through a grid search spanning very small to relatively large cluster granularities, scaled to the number of files in the project.\footnote{The searched ranges and the selected configuration per project are available in our replication package.} We select the configuration that maximizes the Silhouette score among those with a Davies--Bouldin index below 1.0, thereby favouring compact, well-separated clusters. Because these indices measure the quality of a single partition but not its robustness, we additionally assess whether the selected clustering is stable under perturbation. To do so, we re-cluster 30 bootstrap samples of the files using the selected configuration and report the distribution of Adjusted Rand Index (ARI) values against the original clustering.

\subsection{Ownership clustering}
\label{sec:own}
We perform ownership clustering to group files according to their shared developer contributions, based on the intuition that files maintained by the same authors might exhibit similar logging conventions.

\begin{figure}[t]
  \centering
  \includegraphics[width=\textwidth,keepaspectratio,height=4.2cm]{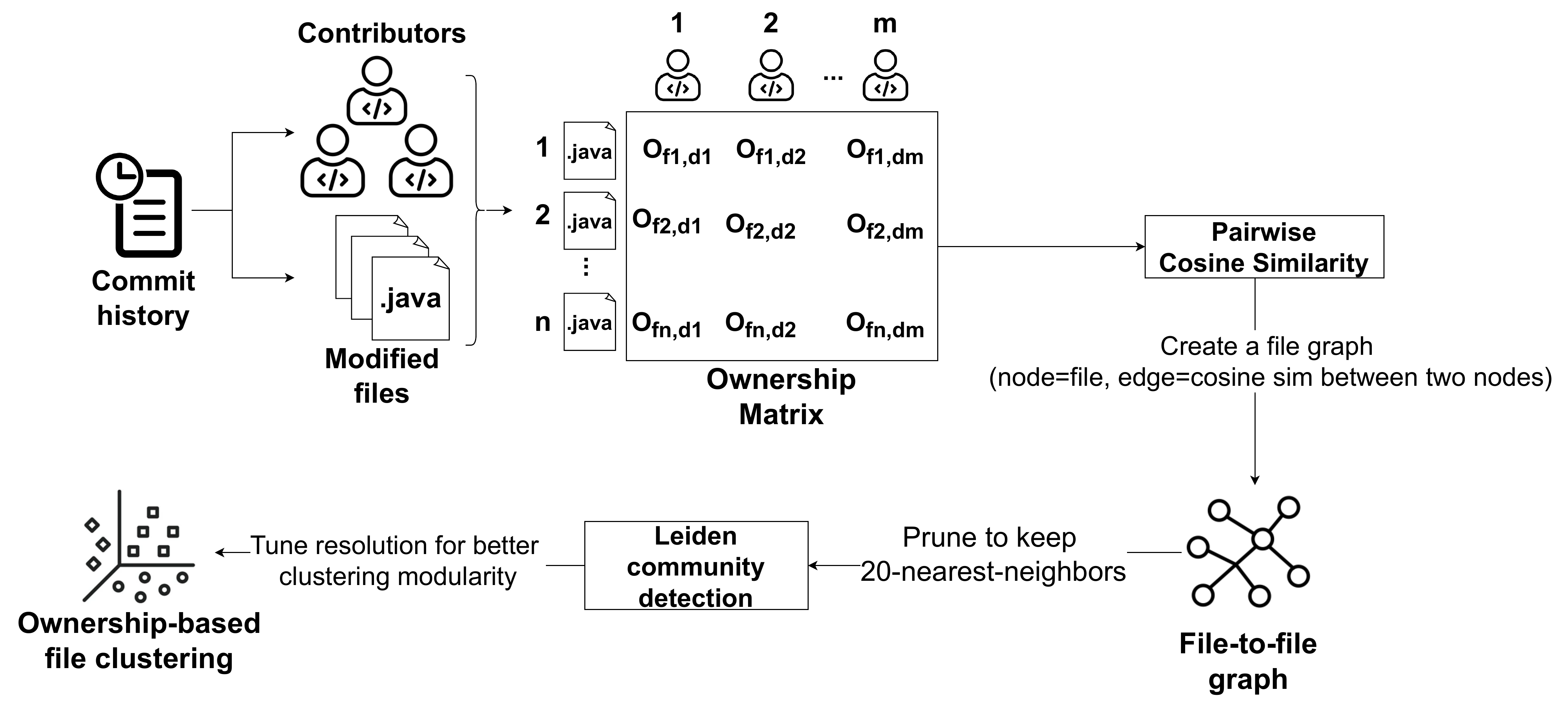}
  \caption{Overview of the ownership-based clustering}
  \label{fig:ownership_clustering}
\end{figure}

\textbf{Raw data construction}: To construct our ownership-based clusters, we mine the Git history of each studied project and build an author–file \emph{ownership matrix} (Figure~\ref{fig:ownership_clustering}). Rows correspond to source files and columns to authors and each cell records how frequently an author has modified a file, with commit counts weighted by an exponential time-decay factor so that recent edits contribute more heavily than older ones. This representation captures both long-term ownership and recent maintenance activity in a single high-dimensional vector for each file.

\textbf{Build a file-to-file graph}: We treat each row of this matrix as a file-level ownership vector and compute pairwise cosine similarities between files. Using these similarities, we construct a weighted file–file graph in which edges connect files that share similar maintainer sets. To avoid an overly dense graph while still preserving strong ownership signals, we connect each file only to its $k$ nearest neighbours by cosine similarity, with $k = 20$. We selected $k=20$ empirically, as lower values fragmented the graph by isolating too many files while higher values merged distinct groups.

\textbf{Build community-based clusters}: We then apply the weighted Leiden community detection algorithm~\cite{Traag:2019} to this $k$-NN graph to obtain ownership communities. Leiden optimizes modularity while avoiding the disconnected communities sometimes produced by the earlier Louvain algorithm~\cite{Blondel:2008}, and scales well to our graph sizes. We tune the resolution parameter and retain settings that yield modularity above 0.7 for all projects, which indicates a pronounced community structure in the ownership graph~\cite{Fortunato:2007,Traag:2019}.

\subsection{Multiplex clustering}
\label{sec:mlpx}
\begin{figure*}[t]
  \centering
  \includegraphics[width=\textwidth,keepaspectratio,height=6.5cm]{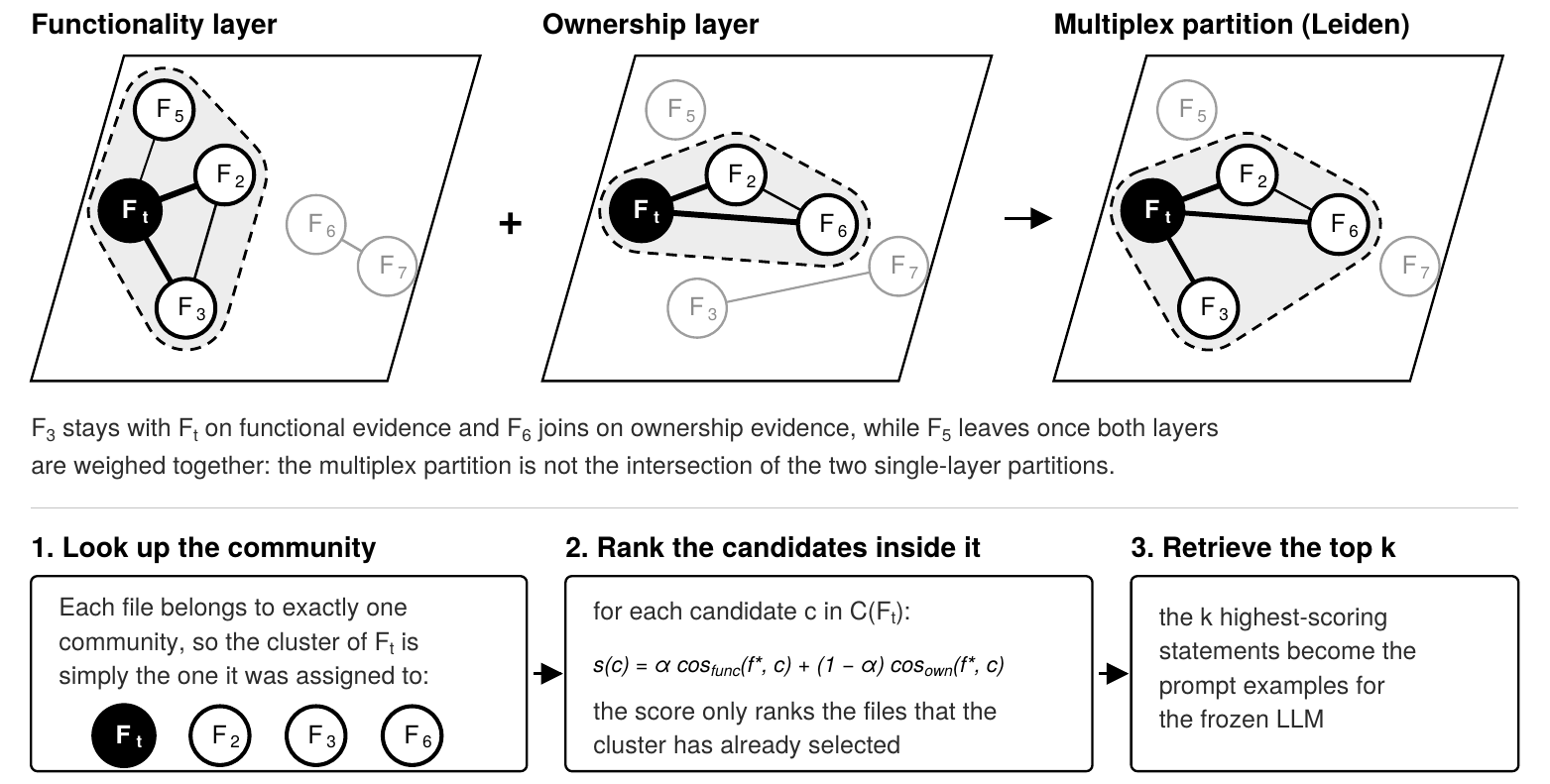}
  \caption{Overview of multiplex clustering and retrieval. Nodes are Java files, present in both layers. The layers differ only in their edges, whose weight reflects functional similarity or co-ownership strength. Dashed outlines mark the community containing the target file $F_t$. The multiplex partition is obtained by optimizing modularity jointly over both layers, and the top-k candidates are then ranked within the resulting community.}
  \label{fig:multiplex_graph}
\end{figure*}

We further investigate the combination of these two socio-technical signals into unified multiplex graph, as shown in Figure~\ref{fig:multiplex_graph}. A multiplex graph is a multi-layered network structure where the same set of nodes exists (i.e., files) across all layers, but each layer represents a distinct type of relationship between them~\cite{Battiston:2014,Domenico:2013}. Specifically, we construct a two-layer multiplex graph, where the first layer encodes \textit{functional similarity} between files based on code embeddings (computed using cosine similarity between file embeddings) and the second layer represents \textit{file ownership} file-to-file graph (Section~\ref{sec:own}) .

Both layers have the same set of nodes (each node corresponding to a Java source file) but maintain distinct edge sets, preserving the unique insights offered by each type of signal. To identify cohesive clusters across these complementary layers, we apply the multiplex variant of the Leiden community detection algorithm~\cite{Traag:2019}. This multiplex algorithm simultaneously optimizes modularity across both the functionality and the ownership layers, so that the resulting communities are cohesive under the combined evidence of the two layers. A resulting community is therefore not the intersection of the two single-layer partitions. For instance, a strong relationship in one layer can keep together files that the other layer alone would have separated.

At inference time, for a given logging statement, our approach identifies the multiplex cluster to which the source file containing the logging statement was assigned, then retrieves the top  $k$ candidate examples from within this cluster. We set  $k=5$ in our experiments as performance flattens beyond this point, with $k=10$ offering no statistically significant improvement at roughly double the prompt length (Appendix Figure~\ref{fig:k_sensitivity}). We then rank the candidates by a combined similarity score
\[
s(c) = \alpha \cdot \cos_{\text{func}}(f^*, c) + (1-\alpha) \cdot \cos_{\text{own}}(f^*, c)
\]
where: 
\begin{itemize}
\item $s(c)$: is the combined similarity score for candidate example $c$. 
\item $f^*$ : is the file containing the logging level to be predicted. 
\item $cos_{func}(f^{*},c)$: is the functional cosine similarity between the embeddings of $f^*$ and candidate $c$. 
\item 
$cos_{own}(f^{*},c)$ is the ownership cosine similarity between the embeddings of $f^*$ and candidate $c$. 
\item $\alpha$ : Weighting coefficient that controls the relative influence of functional similarity versus ownership similarity in the combined retrieval score (tunable with our approach). 
\end{itemize}

Note that $s(c)$ only orders the candidates that the multiplex cluster has already selected. Specifically, the clustering determines which files are eligible, and the two single-layer similarities rank within that set.

We determined the optimal layer weighting $\alpha$ via a grid search on a held-out validation set (a subset of the retrieval data). We evaluated $\alpha$ in increments of 0.1 and selected $\alpha=0.7$, as it maximized the downstream log level prediction AUC score. This value was then fixed for all subsequent experiments on the test sets. In this configuration, our approach gives more influence to functional similarity while still incorporating ownership information. This choice aligns with our design assumption that functional similarity should be the primary retrieval signal, with ownership acting as a secondary refinement. That said $\alpha$ remains a configurable parameter and can be adjusted by practitioners if different trade-offs between functional and ownership signals are desired.

If we cannot identify a suitable multiplex cluster for the target file, our approach applies the following fallbacks in order: (1) functionality cluster retrieval, (2) ownership community retrieval, and (3) random retrieval from the entire project. When a cluster is found but holds fewer than k candidates, all available candidates are used and no fallback is triggered.

\subsection{Prompting LLMs for log level prediction}
To create the log level prediction prompt we leverage the approach explained and detailed in Figure~\ref{fig:prompt_construction} in the appendix. Specifically, for each logging statement we extract the enclosing source code in which it appears, with the logging call itself removed, the text of the log message together with the variables it interpolates, and, for the statements used as examples, the log level its developers assigned to it. All evaluated retrieval strategies share the same fixed prompt template shown in Figure~\ref{fig:prompt_example}, so the only difference between them is which logging examples fill that template.  

\begin{figure}[t]
\centering
\begin{tcolorbox}[colback=gray!5, colframe=black!65, boxrule=0.4pt,
                  left=3pt, right=3pt, top=3pt, bottom=3pt]
\scriptsize\ttfamily\raggedright
You are an experienced software developer. Below are\\
examples of how to choose log levels:\\[3pt]
\textcolor{black!55}{\rmfamily\itshape ---- example 1 of $k$: from \texttt{Threads.java} ----}\\
Instruction: Between trace, debug, info, warn, error,\\
and fatal which is the appropriate log level for this\\
logging statement?\\
Input: The source code is\\
public static void sleep(long millis) \{\\
\hspace*{1em}try \{ Thread.sleep(millis); \}\\
\hspace*{1em}catch (InterruptedException e) \{ \}\\
And the log message is\\
"sleep interrupted", e\\
Response: The log level is \textbf{warn}\\[3pt]
\textcolor{black!55}{\rmfamily\itshape ---- examples $2..k$: same block, decreasing $s(c)$ ----}\\[3pt]
Now do the task:\\[3pt]
Instruction: \textcolor{black!55}{\rmfamily\itshape (as above)}\\
Input: The source code is\\
public void run() \{\\
\hspace*{1em}try \{ mapper.run(subcontext); \}\\
\hspace*{1em}catch (Throwable ie) \{ \}\\
And the log message is\\
"Problem in running map.", ie\\
Response: The log level is \textcolor{black!55}{\rmfamily\itshape <predicted here>}
\end{tcolorbox}
\caption{The prompt template instantiated with logging statements, showing one of the $k$ demonstrations. Grey annotations are not part of the prompt.}
\label{fig:prompt_example}
\end{figure}

For the purposes of this study, we chose four models CodeLlama-7B-Instruct, CodeLlama-34B-Instruct, DeepSeek-V3.2, and GPT-4o, as shown in Table 7 (Appendix). We specifically chose the CodeLlama family (7B and 34B) to isolate the impact of model size on performance within a single, open-source architecture. The 7B version serves as a baseline for consumer-grade hardware, while the 34B version represents the upper limit of what can be deployed on a single consumer-grade GPU node (without quantization). To benchmark these local models against industry standards, we selected GPT-4o as a high-performance commercial baseline, as it is one of the few leading frontier models that provides API access to token log probabilities, which are necessary for our evaluation metrics. Finally, we included DeepSeek-V3.2 to serve as a cost-efficiency benchmark, allowing us to compare our self-hosted approach against a high-performance but low-cost commercial API.

\subsection{Implementation details}
\label{sec:cost}
To validate the practical feasibility of our approach, we measured the execution time of each clustering phase using a representative workstation. This evaluation aims to demonstrate that the socio-technical analysis can be integrated into standard development workflows without imposing prohibitive overhead. In fact, all our presented clustering approaches are computationally efficient, ensuring practical applicability for cluster refreshing if/when needed. Specifically, ownership-based clustering completes in a few minutes, even for our largest studied project (Elasticsearch). Functionality-based clustering takes slightly longer due to the initial step of computing embeddings using a transformer-based tokenizer. However, it remains practical, taking approximately 20 minutes for Elasticsearch (the largest project studied) on standard hardware (16-core CPU, 64 GB RAM, 24 GB GPU memory). Finally, once the artifacts from the prior two clusterings (i.e., ownership matrix and functionality embeddings) are created, multiplex clustering takes few minutes for all of our projects. 

\section{Empirical evaluation setup}
We empirically evaluate the effectiveness of our approach (with its different operation modes) across ten widely used open-source Java projects: Hadoop, HBase, Elasticsearch, Cassandra, Flink, Kafka, JMeter, ZooKeeper, Karaf, and Wicket. These projects were selected due to their varied size, distinct logging practices, and differing clarity in documentation-based component definitions. Specifically, Hadoop and HBase feature explicit documentation-defined components, allowing comparison against a documentation-based retrieval baseline. Conversely, the remaining projects lack clearly defined component boundaries, presenting a scenario ideal for assessing the robustness and generalizability of our approach. Our selection also spans a wide range of logging volumes, from Elasticsearch's 25,187 logging statements down to Wicket's 375, allowing us to assess whether our approach depends on the availability of a large retrieval pool. 

\subsection{Data collection}

We collected the datasets used in this paper by extracting source code files, logging statements, and associated metadata directly from each project's GitHub repository. Specifically, we cloned each repository and checked out the latest stable release versions (e.g., Cassandra-4.1.9) to ensure reproducibility. To extract logging statements systematically, we parsed Java source files using the \textit{javalang}\footnote{https://pypi.org/project/javalang/} library, identifying log invocations and capturing relevant context such as the log message, verbosity level, method boundaries, and surrounding code snippets. Metadata such as file paths, line numbers, and documentation-based components (if any) were also recorded. In cases where a project lacked explicit documentation-based component definitions (e.g., Cassandra), we labeled the component information as \textit{unknown}. 

Table~\ref{tab:project_details} summarizes key characteristics of each of our studied projects.
\begin{table}[t]
\centering
\caption{Summary of the studied projects}
\label{tab:project_details}
\small
\resizebox{\columnwidth}{!}{\begin{tabular}{l l r r r}
\toprule
\textbf{Project} & \textbf{Versions} & \textbf{\# SLOC} & \textbf{\# Contributors} & \textbf{\# Logging statements} \\
\midrule
Hadoop        & 3.3.6 & 1.8M & 635   & 21,910 \\
HBase         & 2.5.9 & 841K & 483   & 10,449 \\
Elasticsearch & 8.9.2 & 2.5M & 2,042 & 25,187 \\
Cassandra     & 4.1.9 & 595K & 464   & 4,808  \\
Flink         & 2.3.0 & 1.7M & 1,671 & 3,800  \\
Kafka         & 4.3.1 & 998K & 1,442 & 3,933  \\
JMeter        & 5.6.3 & 149K & 95    & 1,890  \\
ZooKeeper     & 3.9.5 & 124K & 292   & 1,341  \\
Karaf         & 4.4.11 & 130K & 211  & 665    \\
Wicket        & 10.10.0 & 222K & 163 & 375    \\
\bottomrule
\end{tabular}}
\end{table}

\subsection{Evaluation baselines}
\label{sec:base}
We compare our approach against five baselines: three retrieval strategies and two specialized log level prediction models.
\begin{itemize}
    \item \textbf{Random retrieval (diversity-first)}~\cite{Heng:2024}: in-context-learning examples are randomly selected from the entire dataset of the studied project.

    \item \textbf{Documentation-based retrieval}: for the two projects that define components in their documentation (e.g., HDFS vs. Yarn in Hadoop), examples are retrieved exclusively from within the same documented component, without any clustering. We adopt this strategy for its
    simplicity and its demonstrated efficacy in supervised log level prediction~\cite{Ouatiti:2023}.

    \item \textbf{Flat similarity retrieval}: files across the complete project are ranked by code-embedding cosine similarity (same CodeXEmbed as our approach), and the five examples are drawn from the nearest files.

    \item \textbf{Ordinal Logistic Regression (OLR)}~\cite{Li:2017}: treats log level prediction as an ordinal classification problem over hand-crafted features of the logging statement, its change history, its file and the project history.

    \item \textbf{DeepLV}~\cite{LiZ:2021a}: an ordinal classification approach that encodes the syntactic AST context surrounding the logging statement and its message tokens through an embedding layer and a Bi-LSTM network.
\end{itemize}

We order each project's logging statements chronologically and split them 70\%/30\%, similar to prior work~\cite{Ouatiti:2024}. While the earlier 70\% serves as the retrieval pool for the retrieval-based approaches and as the training set for OLR and DeepLV, the later 30\% is used exclusively for testing. All
retrieval strategies share the same underlying LLM (e.g., CodeLlama-7B), prompting protocol and inference configuration, so that only the mechanism for selecting in-context-learning examples differs.

We evaluate each method on five bootstrap samples of the test set, reporting their mean and standard deviation. Because all models predict the same logging statements, we use paired Wilcoxon signed-rank tests for significance, and report Cohen's $d$ for effect size, interpreting $0.2 \le d < 0.5$ as small, $0.5 \le d < 0.8$ as medium and
$d \ge 0.8$ as large~\cite{Cohen:1992}.

\subsection{Evaluation metrics}
As our approach integrates two modeling tasks (i.e., clustering and log level prediction), we evaluate it using two sets of metrics. To assess clustering quality we use the Adjusted Rand Index (ARI), the Silhouette score, the Davies--Bouldin index (DBI), Density-Based Clustering Validation (DBCV) and Leiden modularity, which we define in the Appendix. To assess log level prediction we use the three metrics established by prior work~\cite{Heng:2024,Li:2017,LiZ:2021a,Ouatiti:2023}: 

\begin{itemize}
    \item \textbf{Area Under the ROC Curve (AUC)}: quantifies the ability of the model to discriminate between log level classes at various threshold settings. To adapt this metric for generative LLMs, we do not rely solely on the discrete text output. Instead, we extract the conditional probabilities of the first generated token corresponding to our target labels (i.e., $\mathcal{V}_{target} = \{\texttt{DEBUG}, \texttt{INFO}, \dots\}$). We normalize these token probabilities to sum to 1 over the candidate set of log levels, effectively treating the LLM as a probabilistic classifier. Following prior work~\cite{Heng:2024,Li:2017,LiZ:2021a,Ouatiti:2023}, we then compute the multiclass AUC using the One-vs-Rest scheme as implemented in scikit-learn\footnote{\url{https://scikit-learn.org/stable/modules/generated/sklearn.metrics.roc_auc_score.html}}.

    \item \textbf{Accuracy}: measures the proportion of logging statements for which the predicted log level exactly matches the ground-truth level, as implemented in scikit-learn~\footnote{\url{https://scikit-learn.org/stable/modules/generated/sklearn.metrics.accuracy_score.html}}. Because accuracy is computed over all statements, it is sensitive to the class imbalance inherent to logging data. We, therefore, report it alongside AUC and AOD rather than in isolation.

    \item \textbf{Average Ordinal Distance Score (AOD)}: captures how close predicted log levels are to their true ordinal positions~\cite{Heng:2024,LiZ:2021a}. Each log level is mapped to an ordinal value (e.g., \texttt{DEBUG} < \texttt{INFO} < \texttt{WARN} < \texttt{ERROR} < \texttt{FATAL}), and AOD normalizes the distance between the predicted and true levels to the $[0,1]$ interval, where higher values indicate predictions that are, on average, closer to the correct log levels. We report AOD as it provides a more nuanced assessment of ordinal prediction quality than accuracy or AUC alone.
\end{itemize}

\section{Results}
\label{sec:results}
\subsection*{\textbf{RQ1. \RQI}}

\textbf{Motivation:} The goal of this research question is to understand the impact of leveraging both functional similarity and developer ownership signals simultaneously to retrieve ICL logging examples for log level prediction scenarios. While the former similarity captures the functional context (i.e., what is being logged), developer ownership captures the human aspect (i.e., who is deciding). Prior research has established that log level selection is influenced by individual developer habits and team norms~\cite{Li:2017, Yuan:2012}. Consequently, retrieving examples based solely on code similarity may yield technically relevant but stylistically inconsistent in-context-learning examples. By integrating the ownership signal with the functionality signal, we align the examples not only with the code's logic but also with the specific logging conventions of the contributor.

\textbf{Approach}: To quantify the impact of multiplex based retrieval (taking into consideration ownership and functionality knowledge simultaneously) on the performance of LLM-powered LLPs, we first perform the multiplex clustering for each studied project, as described in Section~\ref{sec:mlpx}. Then, we evaluate the performance of our approach using in-context-learning examples from the relevant multiplex clusters. We compare this approach to five baselines: (a) random retrieval from the entire project~\cite{Heng:2024}, (b) flat similarity retrieval, (c) documentation-based retrieval when applicable, (d) Ordinal logistic regression LLP~\cite{Li:2017}, and (e) DeepLV (Bi-LSTM)~\cite{LiZ:2021a}. We report in this results section the findings for the Codellama-7B model and refer to the appendix for findings obtained using the Codellama-34B model.

\begin{table*}[t]
\centering
\caption{Summary of the performance of the multiplex retrieval compared to other approaches (metrics are obtained from five bootstrap runs and for CodeLlama-7B)}
\label{tab:main_results}
\resizebox{\textwidth}{!}{\begin{tabular}{l l c c c c c c c c c c}
\toprule
\textbf{Model} & \textbf{Metric} & \textbf{Hadoop} & \textbf{HBase} & \textbf{Elasticsearch} & \textbf{Cassandra} & \textbf{Flink} & \textbf{Kafka} & \textbf{JMeter} & \textbf{ZooKeeper} & \textbf{Karaf} & \textbf{Wicket} \\
\midrule

\multirow{3}{*}{Multiplex retrieval (5)} & AUC       & $\mathbf{0.921 \pm 0.005}$ & $\mathbf{0.914 \pm 0.004}$ & $\mathbf{0.966 \pm 0.002}$ & $\mathbf{0.907 \pm 0.003}$ & $\mathbf{0.875 \pm 0.005}$ & $\mathbf{0.851 \pm 0.006}$ & $\mathbf{0.871 \pm 0.010}$ & $\mathbf{0.866 \pm 0.020}$ & $\mathbf{0.879 \pm 0.041}$ & $\mathbf{0.827 \pm 0.028}$ \\ & Accuracy & $\mathbf{0.694 \pm 0.006}$ & $\mathbf{0.663 \pm 0.015}$ & $\mathbf{0.764 \pm 0.005}$ & $\mathbf{0.668 \pm 0.017}$ & $\mathbf{0.569 \pm 0.008}$ & $\mathbf{0.571 \pm 0.012}$ & $\mathbf{0.607 \pm 0.012}$ & $\mathbf{0.604 \pm 0.032}$ & $0.660 \pm 0.009$ & $\mathbf{0.509 \pm 0.042}$ \\ & AOD       & $\mathbf{0.883 \pm 0.004}$ & $\mathbf{0.868 \pm 0.007}$ & $\mathbf{0.912 \pm 0.002}$ & $\mathbf{0.877 \pm 0.002}$ & $\mathbf{0.844 \pm 0.004}$ & $\mathbf{0.828 \pm 0.007}$ & $0.843 \pm 0.007$ & $\mathbf{0.857 \pm 0.012}$ & $\mathbf{0.868 \pm 0.008}$ & $\mathbf{0.785 \pm 0.018}$ \\ 
\addlinespace

\multirow{3}{*}{Flat similarity retrieval (5)} & AUC & $0.881 \pm 0.003$ & $0.851 \pm 0.002$ & $0.945 \pm 0.002$ & $0.873 \pm 0.012$ & $0.859 \pm 0.003$ & $0.802 \pm 0.009$ & $0.859 \pm 0.009$ & $0.852 \pm 0.010$ & $0.857 \pm 0.072$ & $0.814 \pm 0.026$ \\ & Accuracy & $0.597 \pm 0.009$ & $0.562 \pm 0.011$ & $0.693 \pm 0.003$ & $0.596 \pm 0.020$ & $0.548 \pm 0.006$ & $0.475 \pm 0.011$ & $0.596 \pm 0.022$ & $0.580 \pm 0.016$ & $\mathbf{0.670 \pm 0.017}$ & $0.498 \pm 0.052$ \\ & AOD & $0.844 \pm 0.004$ & $0.826 \pm 0.004$ & $0.885 \pm 0.002$ & $0.849 \pm 0.008$ & $0.837 \pm 0.004$ & $0.784 \pm 0.006$ & $\mathbf{0.844 \pm 0.011}$ & $0.838 \pm 0.008$ & $\mathbf{0.868 \pm 0.011}$ & $0.774 \pm 0.024$ \\ 
\addlinespace

\multirow{3}{*}{Few-random shot (5)~\cite{Heng:2024}}
& AUC & $0.810 \pm 0.002$ & $0.785 \pm 0.005$ & $0.801 \pm 0.003$ & $0.786 \pm 0.017$ & $0.837 \pm 0.005$ & $0.775 \pm 0.002$ & $0.850 \pm 0.006$ & $0.803 \pm 0.007$ & $0.810 \pm 0.019$ & $0.776 \pm 0.005$ \\
& Accuracy & $0.523 \pm 0.003$ & $0.435 \pm 0.005$ & $0.513 \pm 0.016$ & $0.443 \pm 0.014$ & $0.503 \pm 0.013$ & $0.435 \pm 0.007$ & $0.570 \pm 0.016$ & $0.521 \pm 0.018$ & $0.614 \pm 0.035$ & $0.500 \pm 0.018$ \\
& AOD & $0.818 \pm 0.003$ & $0.782 \pm 0.003$ & $0.829 \pm 0.009$ & $0.790 \pm 0.008$ & $0.823 \pm 0.005$ & $0.770 \pm 0.002$ & $0.837 \pm 0.005$ & $0.819 \pm 0.006$ & $0.852 \pm 0.011$ & $0.765 \pm 0.012$ \\
\addlinespace

\multirow{3}{*}{\shortstack[l]{Ordinal Log.\\Regression~\cite{Li:2017}}}
& AUC & $0.731 \pm 0.011$ & $0.729 \pm 0.004$ & $0.766 \pm 0.004$ & $0.710 \pm 0.019$ & $0.677 \pm 0.011$ & $0.628 \pm 0.006$ & $0.682 \pm 0.009$ & $0.653 \pm 0.015$ & $0.723 \pm 0.063$ & $0.643 \pm 0.050$ \\
& Accuracy & $0.359 \pm 0.010$ & $0.367 \pm 0.017$ & $0.329 \pm 0.012$ & $0.371 \pm 0.018$ & $0.292 \pm 0.018$ & $0.272 \pm 0.015$ & $0.407 \pm 0.022$ & $0.294 \pm 0.031$ & $0.359 \pm 0.025$ & $0.293 \pm 0.018$ \\
& AOD & $0.637 \pm 0.007$ & $0.621 \pm 0.007$ & $0.766 \pm 0.004$ & $0.619 \pm 0.012$ & $0.520 \pm 0.010$ & $0.535 \pm 0.008$ & $0.635 \pm 0.017$ & $0.604 \pm 0.023$ & $0.649 \pm 0.021$ & $0.517 \pm 0.036$ \\
\addlinespace

\multirow{3}{*}{DeepLV~\cite{LiZ:2021a}}
& AUC & $0.815 \pm 0.003$ & $0.778 \pm 0.012$ & $0.868 \pm 0.005$ & $0.762 \pm 0.007$ & $0.775 \pm 0.007$ & $0.768 \pm 0.007$ & $0.783 \pm 0.002$ & $0.704 \pm 0.031$ & $0.675 \pm 0.029$ & $0.641 \pm 0.029$ \\
& Accuracy & $0.508 \pm 0.008$ & $0.508 \pm 0.008$ & $0.553 \pm 0.006$ & $0.482 \pm 0.011$ & $0.446 \pm 0.011$ & $0.435 \pm 0.009$ & $0.383 \pm 0.014$ & $0.331 \pm 0.036$ & $0.294 \pm 0.039$ & $0.169 \pm 0.032$ \\
& AOD & $0.790 \pm 0.004$ & $0.752 \pm 0.011$ & $0.807 \pm 0.009$ & $0.757 \pm 0.007$ & $0.763 \pm 0.009$ & $0.701 \pm 0.007$ & $0.763 \pm 0.010$ & $0.774 \pm 0.020$ & $0.650 \pm 0.029$ & $0.620 \pm 0.009$ \\
\bottomrule
\end{tabular}}%
\end{table*}

\textbf{Results}: \textbf{Multiplex retrieval significantly boosts the performance of LLM-powered LLPs, achieving both high AUC and low retrieval latency across all studied projects}, as shown in Table~\ref{tab:main_results}. In fact, leveraging our ownership-functionality aware approach to retrieve in-context-learning logging examples from multiplex clusters results in a statistically significant improvement in AUC (Wilcoxon test, $\alpha = 0.01$, Cohen's d, $d>0.7$), with AUC ranging from 0.83 (Wicket) to 0.97 (Elasticsearch) and exceeding the global random retrieval baseline by a median of 0.07 (0.02 to 0.17 across projects). Similarly, accuracy and AOD exhibit similar statistically significant (Wilcoxon test, $\alpha = 0.01$, Cohen's d, $d>0.7$) improvements as the median accuracy and AOD improves by 0.11 and 0.05 respectively. While flat similarity retrieval also outperforms random retrieval on all ten projects, improving AUC by a median of 0.05 (Wilcoxon test, $\alpha = 0.01$), our multiplex retrieval outperforms flat similarity retrieval on all ten projects by a further median improvement of 0.02 AUC (Wilcoxon test, $\alpha = 0.01$). Thus, while retrieving functionally similar examples provides a substantial improvement over random retrieval, combining functionality and ownership information provides an additional performance benefit. Furthermore, our retrieval step leverages FAISS~\footnote{https://github.com/facebookresearch/faiss} making it highly efficient, with a median latency of 8.9 ms per retrieval, effectively operating at real-time speed (i.e., suitable for IDEs). 
Moreover, our approach statistically significantly (Wilcoxon test, $\alpha = 0.01$, Cohen's d, $d>0.7$) outperforms the established baselines for log level prediction~\cite{Li:2017,LiZ:2021a}.

\textbf{Across our evaluated projects, most mispredictions made by our retrieval consistently involve confusion between adjacent log levels and concentrate in a small subset of files}, as shown in Table~\ref{tab:mispredictions} in the Appendix. Specifically, we observe that adjacent-level mispredictions (e.g., confusing DEBUG with INFO, or WARN with ERROR) account for 54\% (Wicket), to 81\% (Karaf) of total mispredictions. Furthermore, mispredictions are heavily concentrated at the file level, with fewer than 25\% of files accounting for at least half of all mispredictions in each project. This pattern suggests that predictive difficulties faced by our approach are highly localized, providing clear opportunities for targeted improvements. Finally, we note that only a median of 7.3\% of the test logging statements requires a fallback route, either because the socio-technical score cannot be computed or because the cluster holds fewer than  k candidates, and only a median of 0.4\% reaches project-wide random retrieval (Appendix Table~\ref{tab:fallback}).

\begin{tcolorbox}[colback=gray!10, colframe=black, title=\textbf{Summary of RQ1}]
Multiplex clustering, which simultaneously integrates functionality and ownership signals, provides a retrieval strategy that significantly boosts the AUC of LLM-powered LLPs (up to 0.17 improvement over random retrieval). We recommend using our ownership-functionality aware approach for real-time retrieval of high-quality ICL examples supporting log level prediction.
\end{tcolorbox}

\subsection*{\textbf{RQ2. \RQII}}

\textbf{Motivation}: The goal of this research question is to disentangle and quantify the individual contributions of functional similarity and developer ownership signals within our retrieval approach. While both signals offer theoretical utility (functional similarity capturing the context of the code, and ownership capturing the human aspect of the decision making), their comparative effectiveness remains unclear. By isolating these components, we aim to empirically determine whether these signals serve as complementary information channels or if one dimension dominates the predictive capability in log level prediction scenarios.

\textbf{Approach}: To disentangle and quantify the individual contributions of functional similarity and developer ownership signals, we conduct an ablation study by isolating the retrieval mechanisms. Specifically, we instantiate two distinct baselines: (a) Functionality-only retrieval, where examples are selected solely based on vector embedding similarity (as detailed in Section~\ref{sec:sem}), and (b) Ownership-only retrieval, where examples are selected exclusively from the developer's collaboration network (as detailed in Section~\ref{sec:own}). We then evaluate the performance of the LLMs when prompted with examples from these isolated layers and compare them against our integrated ownership-functionality aware approach presented in RQ1. We primarily report the findings for the CodeLlama-7B model in the main text and refer to the appendix for the results obtained using CodeLlama-34B.

\textbf{Results}: \textbf{Our ownership-based clustering approach produces cohesive clusters across all the studied projects}, as shown in Table~\ref{tab:ownership_clustering} (Appendix). In fact, our ownership-based clusters achieve consistently high Leiden modularity scores ($Q$ > 0.74), ranging from 0.74 (ZooKeeper) to 0.93 (Flink). Additionally, intra-cluster cosine similarities (ranging from 0.62 to 0.95) significantly exceed inter-cluster similarities (0.02–0.27), reinforcing the internal cohesion of the identified clusters. Furthermore, the stability analysis of our clustering via the Adjusted Rand Index (ARI) indicates a moderate cluster consistency (median ARI between 0.16 and 0.59), for the six projects with sufficient history. Although Kafka exhibits relatively low stability (ARI of 0.16), its high modularity (0.86) suggests that while ownership may shift frequently, recalculating clusters periodically (e.g., monthly) would effectively sustain cohesive and useful clusters.

Leveraging ownership-based retrieval significantly enhances the predictive performance across nearly all projects, as shown in Table~\ref{tab:full_ownership} (Appendix). Specifically, providing the LLM used for log level prediction with five ownership-based ICL examples statistically significantly (Wilcoxon test, $\alpha$ = 0.01, Cohen’s d, d > 0.8) improves the median AUC of that LLM by 0.02 compared to random retrieval, reaching up to +0.10 on Elasticsearch. Moreover, ownership-based retrieval outperforms (Wilcoxon test, $\alpha= 0.01$, Cohen’s d, d > 0.7) retrieval based on documentation-defined component in one of the two projects for which explicit components are defined in the project’s documentation. The improvements brought by ownership-based retrieval extend to accuracy and AOD scores, for which we report median improvements of 0.04 (accuracy, reaching 0.08) and 0.01 (AOD, reaching 0.03). Our reported scores account for fallback instances where log level predictions are made for files not assigned to any ownership clusters. However, the fallback rate remains minimal (below 10\%) across all studied projects except for HBase, which experiences a higher fallback rate of 33\%. This higher fallback rate indicates that many logging statements fall into files that either belong to small clusters (discarded as noise due to our minimum cluster size constraint of 10 files) or clusters lacking sufficient examples for complete ownership-only retrieval. JMeter is the only project where ownership-only retrieval does not improve on random retrieval in terms of AUC, although the 0.005 deficit is not statistically significant (Wilcoxon test, $\alpha$ = 0.01).

\begin{tcolorbox}[colback=gray!10, colframe=black]
Leveraging file ownership information enhances retrieval quality for LLM-powered LLPs, as ownership-only retrieval achieves a statistically significant improvement in AUC, with a median of 0.02 and reaching 0.10, compared to random retrieval.
\end{tcolorbox}

\textbf{Our functionality-only based clustering yields compact, cohesive, and stable clusters}, as shown in Table~\ref{tab:functionality_clustering} (Appendix). In fact, our clustering approach consistently identifies between 13 and 59 functionally aligned clusters, across our studied projects. These clusters achieve a strong internal cohesion as the median Silhouette score clears the 0.5 cohesion threshold on nine of our ten studied projects (from 0.51 to 0.61), with only ZooKeeper marginally below it at 0.48 while still forming compact clusters (DBCV of 0.37). Across all projects the DBCV ranges between 0.22 and 0.39 (the further from 0 the better). Furthermore, stability of our clusters is high, as demonstrated by the bootstrapped ARI scores ranging from 0.60 (Zookeeper) to 0.84 (Wicket). That said, we observe that HDBSCAN marks 17\% (Karaf) to 49\% (Hadoop) of the files as noise, potentially leading to fallback scenarios during retrieval (i.e., if the log level of the logging statement to be predicted is in a file labeled as noise).

\textbf{Despite the relatively high ratio of files labeled as noise, leveraging functionality-based clusters significantly enhances LLM-powered LLPs compared to our baselines}, as shown in Table~\ref{tab:full_functionality} (Appendix). Indeed, using functionality-only based retrieval to provide in-context-learning logging examples to our LLM-powered LLPs statistically significantly improves the median AUC by 0.05 (up to 0.12) compared to random retrieval, by 0.05 to 0.06 compared to documentation-based in-context-learning retrieval (if any), and by a median of 0.02 (up to 0.06) compared to ownership-based retrieval. The improvement over random retrieval extends to both the accuracy and AOD scores, across all our studied projects. 

\begin{tcolorbox}[colback=gray!10, colframe=black]
Despite labeling a large proportion of files as noise (up to 49\%), functionality-based retrieval statistically significantly enhances the predictive performance of LLM-powered LLPs, improving AUC by a median of 0.05 (up to 0.12) compared to random retrieval. This indicates that code functionality clusters --even with partial coverage-- offer a strong and practical value as a retrieval signal for LLM-powered log level prediction.
\end{tcolorbox}
\subsection*{\textbf{RQ3. \RQIII}}
\textbf{Motivation}: The goal of this research question is to investigate the trade-off between predictive performance and operational efficiency in LLM-powered log level predictors. While commercial, large-scale Large Language Models (e.g., GPT-4o, DeepSeek-V3.2) have established new state-of-the-art benchmarks in code intelligence tasks, their deployment in high-volume software engineering workflows remains constrained by prohibitive inference costs and latency. In fact, log level prediction can be frequently invoked within Continuous Integration (CI) pipelines, where analyzing a large amount of logging statements (along with their context) per commit using a commercial model would incur a large economic overhead. We hypothesize that by enhancing the retrieval mechanism, specifically through our multiplex strategy, we can empower smaller, open-weight models (e.g., 7B parameters) to achieve performance parity with substantially larger commercial models.

\textbf{Approach}: To validate the cost-effectiveness of our approach, we conduct a comparative cost-benefit analysis between our multiplex-enhanced CodeLlama-7B model and two leading commercial models: GPT-4o (OpenAI)~\footnote{https://openai.com/index/hello-gpt-4o/} and DeepSeek-V3.2-chat (DeepSeek)~\footnote{https://api-docs.deepseek.com/news/news251201}. We evaluate these models on the same test sets used in RQ1 and RQ2, recording two key dimensions for each experimental run:

\begin{itemize}
    \item Predictive Performance: Measured via AUC, accuracy and AOD (Similar to RQ1 and RQ2).
    \item Economic Cost: For commercial models, we calculate the estimated monetary cost (in USD) for reproducing the results, based on the official API pricing schemas of the respective providers (e.g., OpenAI) at the time of the experiment. As for our open-weights models we estimate the cost (in USD) based on the hourly rate of the GPU infrastructure (e.g., NVIDIA H100) required to host the model during the inference window. This cost covers inference only. Retrieval preparation is a separate per-project cost, reported in Section~\ref{sec:cost}, and is not included here since its amortized share depends on how often a project's maintainers refresh their clusters.
\end{itemize}

By comparing the commercial LLMs against the 7B model equipped with multiplex retrieval, we aim to determine whether smarter retrieval can substitute for raw model scale. The commercial models are used zero-shot, as five retrieved examples would multiply their input tokens and cost. This contrasts deployment configurations rather than isolating model size, which we assess with CodeLlama-7B and 34B under identical retrieval.

\textbf{Results}: \textbf{Multiplex-enhanced CodeLlama-7B outperforms significantly larger commercial models in terms of AUC across all studied projects}, as shown in Table~\ref{tab:rq3_performance}. 
In fact, we observe that our localized 7B model achieves an average AUC of 0.89, surpassing GPT-4o (0.79) by 0.10 AUC and DeepSeek-V3.2 (0.73) by 0.16 AUC. This performance gap is even more pronounced in the largest datasets, such as Elasticsearch, where the 7B model maintains a high AUC of 0.97 compared to 0.76 for GPT-4o. These results suggest that in the domain of log level prediction, precise context retrieval is a more determinant factor for success than raw parameter scale. By feeding the model project-specific signals via multiplex retrieval, a 7B model can effectively "punch above its weight" and defeat general-purpose models that are estimated to be 20 to 100$\times$ larger.

\begin{table*}[t]
\centering
\caption{Predictive Performance: Local Models vs. Commercial Baselines}
\label{tab:rq3_performance}
\small
\resizebox{\textwidth}{!}{\begin{tabular}{l cc cc cc cc cc cc cc cc cc cc cc}
\toprule
\multirow{2}{*}{\textbf{Model}} & \multicolumn{2}{c}{\textbf{Cassandra}} & \multicolumn{2}{c}{\textbf{HBase}} & \multicolumn{2}{c}{\textbf{Hadoop}} & \multicolumn{2}{c}{\textbf{Elasticsearch}} & \multicolumn{2}{c}{\textbf{Flink}} & \multicolumn{2}{c}{\textbf{Kafka}} & \multicolumn{2}{c}{\textbf{JMeter}} & \multicolumn{2}{c}{\textbf{ZooKeeper}} & \multicolumn{2}{c}{\textbf{Karaf}} & \multicolumn{2}{c}{\textbf{Wicket}} & \multicolumn{2}{c}{\textbf{Average}} \\
\cmidrule(lr){2-3} \cmidrule(lr){4-5} \cmidrule(lr){6-7} \cmidrule(lr){8-9} \cmidrule(lr){10-11} \cmidrule(lr){12-13} \cmidrule(lr){14-15} \cmidrule(lr){16-17} \cmidrule(lr){18-19} \cmidrule(lr){20-21} \cmidrule(lr){22-23}
 & \textbf{AUC} & \textbf{AOD} & \textbf{AUC} & \textbf{AOD} & \textbf{AUC} & \textbf{AOD} & \textbf{AUC} & \textbf{AOD} & \textbf{AUC} & \textbf{AOD} & \textbf{AUC} & \textbf{AOD} & \textbf{AUC} & \textbf{AOD} & \textbf{AUC} & \textbf{AOD} & \textbf{AUC} & \textbf{AOD} & \textbf{AUC} & \textbf{AOD} & \textbf{AUC} & \textbf{AOD} \\
\midrule
\textbf{CodeLlama-34B (Multiplex)} & \textbf{0.914} & \textbf{0.884} & \textbf{0.926} & \textbf{0.880} & \textbf{0.932} & \textbf{0.894} & \textbf{0.969} & \textbf{0.913} & \textbf{0.887} & \textbf{0.850} & \textbf{0.860} & 0.826 & \textbf{0.877} & \textbf{0.857} & \textbf{0.879} & 0.840 & \textbf{0.879} & \textbf{0.873} & \textbf{0.836} & \textbf{0.811} & \textbf{0.896} & \textbf{0.863} \\
CodeLlama-7B (Multiplex) & 0.907 & 0.877 & 0.914 & 0.868 & 0.921 & 0.883 & 0.966 & 0.912 & 0.875 & 0.844 & 0.851 & \textbf{0.828} & 0.871 & 0.843 & 0.866 & \textbf{0.857} & \textbf{0.879} & 0.868 & 0.827 & 0.785 & 0.888 & 0.857 \\
GPT-4o (Zero-shot) & 0.806 & 0.798 & 0.776 & 0.807 & 0.809 & 0.814 & 0.760 & 0.780 & 0.832 & 0.828 & 0.765 & 0.766 & 0.858 & 0.843 & 0.792 & 0.836 & 0.759 & 0.857 & 0.755 & 0.708 & 0.791 & 0.804 \\
DeepSeek-V3.2 (Zero-shot) & 0.669 & 0.779 & 0.693 & 0.802 & 0.679 & 0.794 & 0.595 & 0.775 & 0.807 & 0.823 & 0.746 & 0.763 & 0.840 & 0.835 & 0.788 & 0.815 & 0.749 & 0.850 & 0.741 & 0.769 & 0.731 & 0.801 \\
\bottomrule
\end{tabular}}
\end{table*}

\textbf{The CodeLlama-7B model offers the best trade-off between predictive performance, data privacy, and operational cost}, as shown in Table~\ref{tab:cost_analysis}. While DeepSeek-V3.2 offers the lowest absolute price for sequential processing ($\approx$\$0.07 per 1k logs), this saving imposes a large penalty of 18\% in predictive AUC and requires the transmission of proprietary code to external servers. In contrast, our self-hosted 7B model maintains superior performance (0.89 AUC) and strict data sovereignty. Furthermore, by leveraging concurrent batching on our experimental hardware, we achieve a unit cost of \$0.18 per 1k logs, as shown in Table~\ref{tab:cost_analysis}. While slightly higher than the cheapest API, this cost remains negligible for high-value CI pipelines while mitigating the risks of data leakage and external service outages.


\begin{table}[t]
\centering
\caption{Operational Cost Comparison: Self-Hosted vs. Commercial APIs (per 1k logs)}
\label{tab:cost_analysis}
\small
\resizebox{\columnwidth}{!}{%
\begin{tabular}{l l r c}
\toprule
\textbf{Model Strategy} & \textbf{Inference Method} & \textbf{Cost} & \textbf{vs. DeepSeek} \\
\midrule
\textbf{CodeLlama-7B} & Local (Batched) & \textbf{\$0.18} & \textbf{Competitive} \\
CodeLlama-34B & Local (Batched) & \$0.76 & +1,051\% \\
\midrule
DeepSeek-V3.2 & API (Token Rate) & \$0.066 & Reference \\
GPT-4o & API (Token Rate) & \$2.44 & +3,598\% \\
\bottomrule
\end{tabular}%
}
\end{table}
\textbf{Scaling up the local model from 7B to 34B parameters yields diminishing returns}, as shown in Table~\ref{tab:rq3_scaling}. In our standardized experimental setup, we observe that while the CodeLlama-34B model achieves the highest absolute AUC of 0.90, it offers only a marginal improvement of 0.9\% over the 7B model. However, this increment comes at a steep price, as the 34B model is significantly more computationally intensive, slowing down batch throughput by 4.1$\times$. This throughput degradation directly translates to a 322\% increase in operational costs (rising from \$0.18 to \$0.76 per 1k logs). Consequently, for CI/CD integrated log level prediction (where throughput is critical), the 7B model represents the superior efficiency choice.

\begin{table}[h]
\centering
\caption{The Scaling Analysis: Performance Gains vs. Operational Cost Increases (4x H100)}
\label{tab:rq3_scaling}
\small
\resizebox{\columnwidth}{!}{%
\begin{tabular}{l c c c c c}
\toprule
\textbf{Model Size} &  \textbf{AUC} & \textbf{\begin{tabular}[c]{@{}c@{}}Cost per\\1k Logs\end{tabular}} & \textbf{\begin{tabular}[c]{@{}c@{}}Performance\\Gain (AUC)\end{tabular}} & \textbf{\begin{tabular}[c]{@{}c@{}}Cost\\Increase\end{tabular}} \\
\midrule
7B (Base) &  0.89 & \textbf{\$0.18} & -- & -- \\
34B (Large) &  \textbf{0.90} & \$0.76 & +0.9\% & +322\% \\
\bottomrule
\end{tabular}%
}
\end{table}

\begin{tcolorbox}[colback=gray!10, colframe=black, title=\textbf{Summary of RQ3}]
Our results show that a multiplex-retrieval enhanced 7B model outperforms commercial frontier models (GPT-4o, DeepSeek-V3.2) by up to 0.16 in terms of AUC for log level prediction, suggesting that our specialized retrieval strategy is more critical than raw parameter scale. While the budget API (i.e., Deepseek) appears cheaper for low volumes, batched local inference achieves competitive unit costs ($\approx$\$0.18/1k logs) without data transmission risks. We recommend practitioners prioritize localized, retrieval-augmented models to ensure data sovereignty and reliable log level recommendations.
\end{tcolorbox}

\section{Threats to validity}

\subsection{Internal Validity}
An internal validity threat concerns our reliance on embedding models to represent files for clustering and retrieval. While we employed a state-of-the-art embedding model (CodeXEmbed) due to its proven effectiveness in capturing code semantics, any inherent biases or limitations in this model could influence the quality of our clustering outcomes, consequently affecting the performance of LLM-powered LLPs using those clusters. To mitigate this threat, we validated cluster quality using established metrics (Silhouette, DBCV, and modularity) and assessed their stability through bootstrapping and temporal re-clustering analyses. Future work could benefit from exploring different embedding approaches to further strengthen the generalizability and robustness of the results.

\subsection{Construct Validity}
A construct validity threat involves the selection of clustering hyperparameters, such as the resolution parameter ($\gamma$) and cluster size thresholds (e.g., minimum cluster size). Arbitrary or suboptimal parameter choices could impact clustering results. To address this concern, we conducted specialized grid searches for each of our projects, identifying optimal parameter values based on robust clustering quality evaluation metrics, therefore ensuring our results' consistency and reducing sensitivity to hyperparameter choices.

Another construct validity threat concerns what our metrics measure. We evaluate how well a predictor recovers the log level that a project's maintainers assigned to a logging statement, which we treat as a reference label rather than as a certified optimum. To reduce the risk that these labels capture transient rather than settled decisions, we take levels from stable releases, following prior work~\cite{Li:2017,LiZ:2021a,Heng:2024}. Logging statements are frequently revised after being introduced~\cite{Kabinna:2016b}, so levels observed at a stable release are more likely to reflect settled maintainer judgment than levels sampled from arbitrary commits. That said, agreement with maintainer-assigned levels does not establish that predicted levels improve debugging effectiveness or system observability. Establishing that would require a different study design, for instance measuring developer acceptance and override rates on real change requests, or comparing the diagnostic coverage of a system instrumented with predicted versus maintainer-assigned levels. Our results should therefore be read as evidence about the quality of the recommendations a log level predictor produces. We consider this a useful measure, given that inconsistent log levels are reported to be a source of maintenance effort.

\subsection{External Validity}
One external validity threat is our focus on ten large open-source Java projects, as findings may not generalize to smaller, proprietary, or non-Java systems. This applies in particular to our ownership signal, since ownership in these projects emerges from voluntary contribution patterns, whereas industrial systems may assign maintenance responsibility through organizational policy and produce different ownership structures. Future work could replicate our analysis across diverse software ecosystems and languages.

Another external validity threat relates to the LLMs used in our study. While our evaluation includes both open-weight (CodeLlama-7B, 34B) and commercial state-of-the-art models (GPT-4o, DeepSeek-V3.2), our investigation into local deployment focused primarily on the CodeLlama family. Although CodeLlama is a widely adopted standard for code tasks, different open-weight architectures may exhibit distinct behaviors when coupled with our multiplex retrieval strategy. A related threat is that these models were pretrained on public code and may already encode logging patterns from the studied projects. Because every retrieval strategy we compare queries the same frozen model with the same prompt and test statements, any such exposure is held constant across conditions and the reported differences remain attributable to retrieval. We cannot, however, rule out an interaction in which retrieval helps more or less on memorized code, and absolute performance on private code may differ. We mitigate this by leveraging multiple robust evaluation metrics (AUC, accuracy, AOD) and conducting statistical analyses of performance improvements. Future studies could explore different LLM architectures and their impact on log level prediction.

One final external validity threat relates to our approach's reliance on two specific clustering signals (functional similarity and developer ownership). Other signals --such as defect density, co-change coupling, package or module hierarchy, or call-graph proximity-- might also influence developers' log level decisions and could lead to further performance gains. Our design requires that each signal be derivable consistently at file granularity and representable as an aligned layer of the same multiplex graph, a constraint that not all candidate signals satisfy. While our results show that these two signals yield significant predictive performance improvements, future research could explore incorporating additional signals to further enhance our approach's predictive capabilities.
\section{Conclusion}
Selecting appropriate log levels for new logging statements is an important, yet challenging software engineering task. Prior studies have leveraged machine learning models to automate log level prediction. However, these approaches have either focused on local, code-driven metrics (like AST context) or used file-wide engineered features, overlooking the socio-technical knowledge of a project.

In this paper, we conducted a comparative study to investigate the impact of these signals: code functionality and developer ownership. We hypothesized that a modern LLM, guided by a novel retrieval strategy, could outperform classic models and commercial models with hundreds of billions of parameters. To test our hypothesis, we introduced ownership-functionality aware retrieval, an in-context-learning (ICL) retrieval approach that integrates functionality and ownership information simultaneously through multiplex clustering.

Our evaluation across ten popular open-source projects demonstrates the superiority of this new paradigm. We show that leveraging these combined signals for in-context-learning retrieval establishes a new state-of-the-art for LLM-powered log level prediction, achieving a median AUC between 0.83 and 0.97. This result significantly outperforms the naive LLM baseline and prior specialized supervised learning paradigms, including deep learning (DeepLV: Avg. AUC 0.76) and feature-engineered Ordinal Regression (Avg. AUC 0.69).

Our findings provide evidence that leveraging our evaluated socio-technical signals (i.e., code functionality and ownership) can have a high impact on the predictive performance of LLM-powered LLPs. We envision this retrieval-augmented approach combining both signals simultaneously being extended to other logging tasks, such as log message generation and placement recommendations, further assisting developers in producing high quality logs.

\bibliographystyle{ieeetrs}
\bibliography{loganalysis_short}
\clearpage
\section*{Appendix}
\label{sec:appendix}

\subsection*{Prompting LLMs for log level prediction}
\begin{figure*}[!htbp]
  \centering
  \includegraphics[width=\textwidth,keepaspectratio,height=7cm]{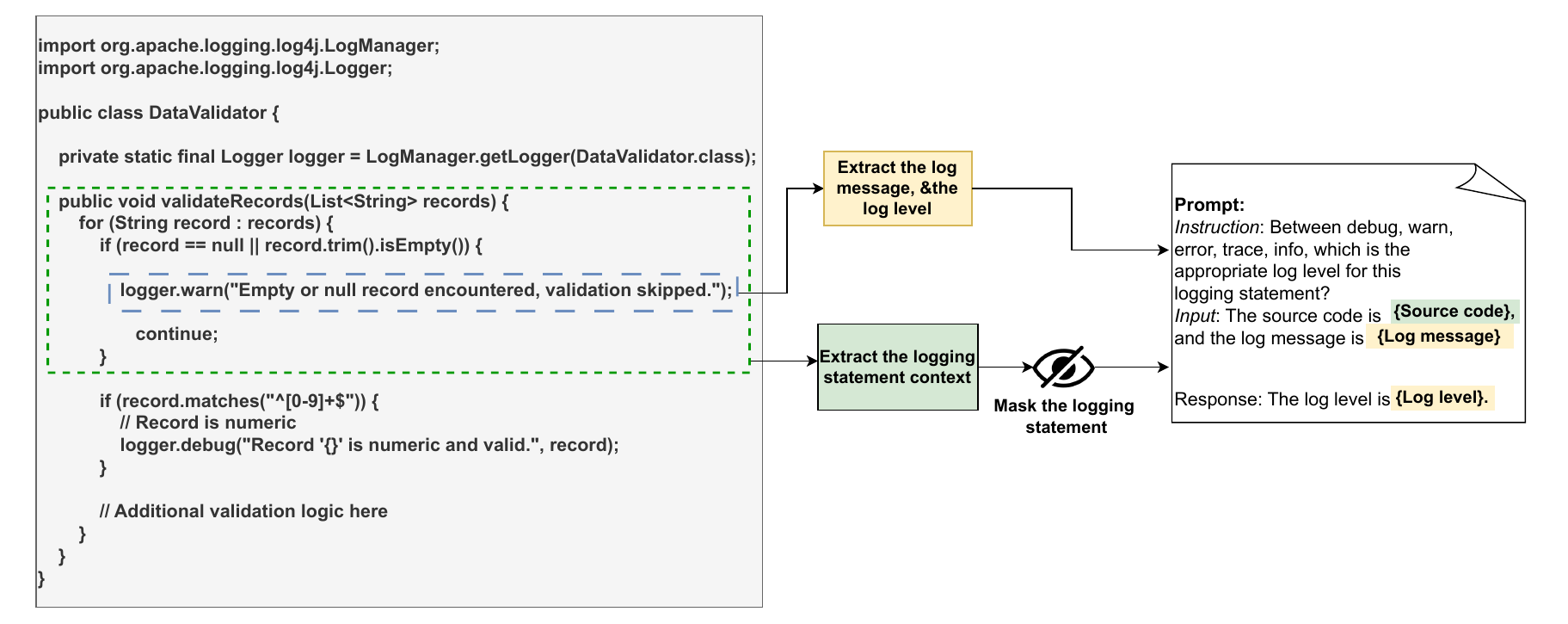}
  \caption{Prompt creation process for the LLM-powered log level prediction. The candidate log levels listed in the instruction are those the project actually uses, and therefore vary across projects.}
  \label{fig:prompt_construction}
\end{figure*}

To create the log level prediction prompt we leverage the approach detailed in Figure~\ref{fig:prompt_construction}. Specifically, we first parse the code to isolate the target logging instruction (e.g., \texttt{logger.warn(...)}) from its surrounding method. The process involves three key steps:
\begin{itemize}
    \item \textbf{Extraction:} We separate the log message (the text identifying the event) and the ground truth log level from the source code.
    \item \textbf{Masking:} We extract the surrounding code context to provide the model with syntactic cues. Crucially, we mask the actual logging statement within this context. This prevents data leakage and forces the model to infer the level based on the logic of the code rather than simply reading the existing statement (e.g., \texttt{if (logger.isDebugEnabled())}).
    \item \textbf{Prompt Assembly:} The final prompt is constructed by combining a specific task instruction (defining the valid output format), the masked source code context, and the isolated log message.
\end{itemize}

\subsection*{LLMs used in this study}
For the purposes of our study, we leverage the LLMs shown in Table~\ref{tab:models}.

\begin{table*}[t]
\centering
\caption{Summary of Large Language Models Selected for Evaluation}
\label{tab:models}
\small
\begin{tabular}{l l l c c c}
\toprule
\textbf{Model} & \textbf{Provider} & \textbf{Architecture} & \textbf{Parameters} & \textbf{Context} & \textbf{Access} \\
& & & \textit{(Total / Active)} & \textit{(Window)} & \\
\midrule
\multicolumn{6}{l}{\textit{Open-Weight Models}} \\
\midrule
CodeLlama-7B-Instruct & Meta & Dense Decoder & 7B / 7B & 16k & Open Weights \\
CodeLlama-34b-Instruct & Meta & Dense Decoder & 34B / 34B & 16k & Open Weights \\
DeepSeek-V3.2 & DeepSeek & MoE (Sparse Attn) & 685B / 37B & 164k & Open Weights \\
\midrule
\multicolumn{6}{l}{\textit{Proprietary Models}} \\
\midrule
GPT-4o & OpenAI & Undisclosed\textsuperscript{*} & Unknown & 128k & API \\
\bottomrule
\end{tabular}
\vspace{1ex}
\begin{minipage}{\linewidth}
\footnotesize
\textit{Note:} For proprietary models, exact architectural details and parameter counts are not public.
\textsuperscript{*}While technically undisclosed, community analysis suggests GPT-4o utilizes a Mixture-of-Experts (MoE) architecture.
\end{minipage}
\end{table*}

\subsection*{Clustering quality metrics}
We assess the quality of our functionality-based and ownership-based clusterings (Tables~\ref{tab:ownership_clustering} and~\ref{tab:functionality_clustering}) using the following metrics:
\begin{itemize}
    \item Adjusted Rand Index~\footnote{\url{https://scikit-learn.org/stable/modules/generated/sklearn.metrics.adjusted_rand_score.html}} (ARI): quantifies the agreement or similarity between two clustering solutions. ARI ranges between -1 and 1, where values close to 1 indicate strong agreement, values close to 0 indicate random clustering, and negative values indicate worse-than-random clustering~\cite{Steinley:2004}.
    \item Silhouette score~\footnote{\url{https://scikit-learn.org/stable/modules/generated/sklearn.metrics.silhouette_score.html}}: measures how tightly grouped (i.e., cohesive) points within clusters are compared to points in other clusters. It ranges from -1 to 1, with values above 0.5 indicate strong, cohesive, and well-separated clusters; values between 0.25 and 0.5 reflect moderate clustering structure; and values below 0.25 suggest weak or ambiguous clustering~\cite{Rousseeuw:1987}. 
    \item Davies–Bouldin index~\cite{Davies:1979} (DBI): measures how well clusters are separated by comparing each cluster’s internal dispersion with its nearest-neighbour distance. It ranges from 0 to $\infty$, where lower values are better. Typically, DBI scores below 1 indicate well-separated and compact clusters~\cite{Davies:1979}. 
    \item     Density-Based Clustering Validation (DBCV)~\cite{Moulavi:2014}: a density-aware validation metric tailored for algorithms such as HDBSCAN. DBCV combines cluster compactness (how tightly points are packed around local density modes) with cluster isolation (how deep the density valleys are between neighbouring clusters). Values range from $-1$ to $+1$, with positive scores indicating well-formed, well-separated density regions, values near 0 suggesting structures that are indistinguishable from noise, and negative scores indicating badly fragmented or overlapping clusters.
    \item  Leiden modularity~\cite{Traag:2019}: assesses how a graph can be partitioned into distinct clusters by providing insight into the overall strength and quality of cluster boundaries. High modularity indicates that nodes within clusters have dense internal connections and relatively sparse connections to nodes in other clusters. Values above 0.3 typically indicate a strong and meaningful community structure~\cite{Fortunato:2007,Traag:2019} .

\end{itemize}

\subsection*{Retrieval step coverage}
\label{sec:appendix_coverage}
Table~\ref{tab:fallback} reports, for each studied project, the percentage of test logging statements served by each of the three retrieval steps described in Section~\ref{sec:mlpx}, together with a breakdown of the routes taken within the multiplex step.

\begin{table*}[t]
\centering
\caption{Coverage of each retrieval step, per project (CodeLlama-7B, five in-context-learning examples). Percentages are computed once over the unique test logging statements of each project. $\dagger$:the file that contains the statement lacks one of the two signals, so one of the two cosine similarities is undefined and the examples are drawn uniformly from the same multiplex cluster. $\ddagger$: the cluster holds fewer than $k$ candidates, so all of them are used. Step 2 (single-signal retrieval) is never invoked across our studied projects, as the multiplex assignment fully subsumes what functionality-only or ownership-only retrieval would have covered.}
\label{tab:fallback}
\resizebox{\textwidth}{!}{\begin{tabular}{l c c c c c c c c c c}
\toprule
& \textbf{Hadoop} & \textbf{HBase} & \textbf{Elasticsearch} & \textbf{Cassandra} & \textbf{Flink} & \textbf{Kafka} & \textbf{JMeter} & \textbf{ZooKeeper} & \textbf{Karaf} & \textbf{Wicket} \\
\midrule
\textbf{Step 1: multiplex cluster retrieval} & \textbf{100.0\%} & \textbf{100.0\%} & \textbf{99.4\%} & \textbf{99.8\%} & \textbf{95.1\%} & \textbf{99.1\%} & \textbf{100.0\%} & \textbf{100.0\%} & \textbf{91.7\%} & \textbf{99.1\%} \\
\quad Ranked by the socio-technical score & 90.7\% & 67.3\% & 97.3\% & 95.3\% & 85.8\% & 95.0\% & 95.9\% & 93.8\% & 91.7\% & 64.3\% \\
\quad Sampled from the multiplex cluster$^{\dagger}$ & 9.1\% & 32.7\% & 0.5\% & 4.0\% & 1.4\% & 0.9\% & 0.9\% & 6.2\% & 0.0\% & 0.0\% \\
\quad Fewer than $k$ candidates in the cluster$^{\ddagger}$ & 0.2\% & 0.0\% & 1.7\% & 0.5\% & 7.9\% & 3.2\% & 3.2\% & 0.0\% & 0.0\% & 34.8\% \\
\midrule
\textbf{Step 2: single-signal retrieval} & \textbf{0.0\%} & \textbf{0.0\%} & \textbf{0.0\%} & \textbf{0.0\%} & \textbf{0.0\%} & \textbf{0.0\%} & \textbf{0.0\%} & \textbf{0.0\%} & \textbf{0.0\%} & \textbf{0.0\%} \\
\textbf{Step 3: project-wide random retrieval} & \textbf{0.0\%} & \textbf{0.0\%} & \textbf{0.6\%} & \textbf{0.2\%} & \textbf{4.9\%} & \textbf{0.9\%} & \textbf{0.0\%} & \textbf{0.0\%} & \textbf{8.3\%} & \textbf{0.9\%} \\
\bottomrule
\end{tabular}}%
\end{table*}

\subsection*{Misprediction analysis}
\label{sec:appendix_mispredictions}
Table~\ref{tab:mispredictions} characterizes the mispredictions made by
multiplex retrieval. Adjacent-level mispredictions (e.g., confusing
\texttt{DEBUG} with \texttt{INFO}, or \texttt{WARN} with \texttt{ERROR})
account for 54\% (Wicket) to 81\% (Karaf) of all mispredictions, and fewer
than 25\% of files account for at least half of the mispredictions in each
project.

\begin{table}[t]
\centering
\caption{Overview of the mispredictions using multiplex retrieval}
\label{tab:mispredictions}
\small
\resizebox{\columnwidth}{!}{\begin{tabular}{l c c}
\toprule
\textbf{Project} & \textbf{\begin{tabular}[c]{@{}c@{}}Adjacent Level\\Confusion\end{tabular}} & \textbf{\begin{tabular}[c]{@{}c@{}}Files Covering 50\%\\of Mispredictions\end{tabular}} \\
\midrule
Cassandra     & 64.1\% & 13.3\% \\
Elasticsearch & 64.7\% & 16.5\% \\
Flink         & 78.6\% & 21.8\% \\
Hadoop        & 72.8\% & 14.7\% \\
HBase         & 67.0\% & 13.9\% \\
JMeter        & 73.4\% & 19.8\% \\
Kafka         & 66.5\% & 16.8\% \\
Karaf         & 81.3\% & 24.2\% \\
Wicket        & 53.7\% & 23.8\% \\
ZooKeeper     & 80.4\% & 17.0\% \\
\bottomrule
\end{tabular}}
\end{table}

\subsection*{RQ2: Single signal retrievals (Ablation study)}
\label{sec:appendix_rq2}
\begin{figure}[t]
\centering
\includegraphics[width=\columnwidth]{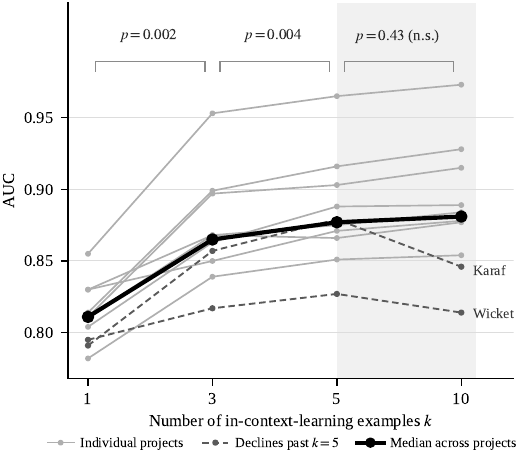}
\caption{Sensitivity of multiplex retrieval to the number of in-context-learning examples $k$ (AUC, CodeLlama-7B, five bootstrap runs). Grey lines are individual projects; the black line is the median across the ten studied projects. Gains are statistically significant up to five examples and flatten beyond it.}
\label{fig:k_sensitivity}
\end{figure}

We first report the clustering quality metrics of our ownership and functionality clustering in Tables~\ref{tab:ownership_clustering} and~\ref{tab:functionality_clustering}. We then report the full performance results for CodeLlama-7B ownership-only (Table~\ref{tab:full_ownership}) and functionality-only (Table~\ref{tab:full_functionality}) retrieval. Finally, Table~\ref{tab:appendix_34b_ablation} reports the impact of the different retrieval strategies on the CodeLlama-34B performance, and Figure~\ref{fig:k_sensitivity} reports the sensitivity of multiplex retrieval to the number of in-context-learning examples.

\begin{table}[H]
\centering
\caption{Ownership-based clustering quality metrics. A dash indicates too few non-empty two-month windows for a reliable stability estimate.}
\label{tab:ownership_clustering}
\small
\resizebox{\columnwidth}{!}{\begin{tabular}{l c c c r}
\toprule
\textbf{Project} & \textbf{Modularity ($Q_c$)} & \textbf{Median ARI} & \textbf{\begin{tabular}[c]{@{}c@{}}Intra/Inter \\ Cosine Similarity\end{tabular}} & \textbf{Coverage} \\
\midrule
Hadoop        & 0.798 & 0.524 & 0.851 / 0.123 & 99.0\% \\
HBase         & 0.805 & 0.286 & 0.618 / 0.018 & 67.0\% \\
Elasticsearch & 0.850 & 0.556 & 0.720 / 0.268 & 99.5\% \\
Cassandra     & 0.919 & 0.588 & 0.651 / 0.093 & 96.0\% \\
Flink         & 0.933 & 0.447 & 0.733 / 0.037 & 95.9\% \\
Kafka         & 0.862 & 0.158 & 0.630 / 0.032 & 95.9\% \\
JMeter        & 0.788 & --    & 0.945 / 0.265 & 99.1\% \\
ZooKeeper     & 0.743 & --    & 0.799 / 0.111 & 93.8\% \\
Karaf         & 0.829 & --    & 0.899 / 0.170 & 100.0\% \\
Wicket        & 0.901 & --    & 0.876 / 0.150 & 100.0\% \\
\bottomrule
\end{tabular}}
\end{table}

\begin{table}[H]
\centering
\caption{Functionality-based clustering quality metrics}
\label{tab:functionality_clustering}
\small
\resizebox{\columnwidth}{!}{\begin{tabular}{l r c c c c c c}
\toprule
\textbf{Project} & \textbf{Files} & \textbf{\begin{tabular}[c]{@{}c@{}}Best Param.\\(mcs / ms)\end{tabular}} & \textbf{\# Clusters} & \textbf{Silh.} & \textbf{DBCV} & \textbf{Noise} & \textbf{\begin{tabular}[c]{@{}c@{}}Bootstr.\\ARI\end{tabular}} \\
\midrule
Hadoop        & 7,395  & 50 / 25  & 31 & 0.533 & 0.230 & 49\% & $0.685 \pm 0.027$ \\
HBase         & 2,691  & 25 / 12  & 27 & 0.566 & 0.325 & 41\% & $0.690 \pm 0.040$ \\
Elasticsearch & 14,976 & 100 / 25 & 31 & 0.553 & 0.219 & 45\% & $0.670 \pm 0.050$ \\
Cassandra     & 3,158  & 25 / 12  & 30 & 0.606 & 0.347 & 37\% & $0.700 \pm 0.030$ \\
Flink         & 14,691 & 50 / 25  & 59 & 0.561 & 0.330 & 39\% & $0.806 \pm 0.025$ \\
Kafka         & 5,864  & 50 / 25  & 22 & 0.541 & 0.387 & 30\% & $0.797$ \\
JMeter        & 1,402  & 26 / 13  & 16 & 0.526 & 0.349 & 31\% & $0.674$ \\
ZooKeeper     & 951    & 18 / 5   & 16 & 0.478 & 0.366 & 28\% & $0.604 \pm 0.137$ \\
Karaf         & 1,631  & 30 / 7   & 16 & 0.505 & 0.394 & 17\% & $0.782 \pm 0.065$ \\
Wicket        & 3,324  & 50 / 25  & 13 & 0.592 & 0.303 & 27\% & $0.840$ \\
\bottomrule
\end{tabular}}
\end{table}

\begin{table*}[t]
\centering
\caption{Summary of the performance of the ownership-only retrieval compared to other approaches (metrics are obtained from 5 bootstrap runs and for CodeLlama-7B). Documentation-based retrieval is only applicable to the two projects that define components in their documentation; a dash indicates the configuration is not applicable for that project.}
\label{tab:full_ownership}
\small
\resizebox{\textwidth}{!}{\begin{tabular}{l l c c c c c c c c c c}
\toprule
\textbf{Model} & \textbf{Metric} & \textbf{Hadoop} & \textbf{HBase} & \textbf{Elasticsearch} & \textbf{Cassandra} & \textbf{Flink} & \textbf{Kafka} & \textbf{JMeter} & \textbf{ZooKeeper} & \textbf{Karaf} & \textbf{Wicket} \\
\midrule

\multirow{3}{*}{\textbf{Ownership-based (5)}}
& AUC & $\mathbf{0.832 \pm 0.002}$ & $0.808 \pm 0.002$ & $\mathbf{0.904 \pm 0.001}$ & $\mathbf{0.837 \pm 0.007}$ & $\mathbf{0.855 \pm 0.006}$ & $\mathbf{0.785 \pm 0.009}$ & $0.845 \pm 0.012$ & $\mathbf{0.831 \pm 0.029}$ & $\mathbf{0.833 \pm 0.061}$ & $\mathbf{0.789 \pm 0.021}$ \\
& Accuracy & $\mathbf{0.532 \pm 0.002}$ & $\mathbf{0.510 \pm 0.008}$ & $\mathbf{0.578 \pm 0.006}$ & $\mathbf{0.501 \pm 0.019}$ & $\mathbf{0.521 \pm 0.015}$ & $\mathbf{0.461 \pm 0.008}$ & $0.531 \pm 0.012$ & $\mathbf{0.571 \pm 0.028}$ & $0.605 \pm 0.029$ & $\mathbf{0.554 \pm 0.051}$ \\
& AOD & $\mathbf{0.819 \pm 0.001}$ & $\mathbf{0.808 \pm 0.002}$ & $\mathbf{0.845 \pm 0.001}$ & $\mathbf{0.824 \pm 0.008}$ & $\mathbf{0.825 \pm 0.008}$ & $\mathbf{0.778 \pm 0.004}$ & $0.826 \pm 0.009$ & $\mathbf{0.832 \pm 0.015}$ & $0.850 \pm 0.011$ & $\mathbf{0.796 \pm 0.029}$ \\
\addlinespace

\multirow{3}{*}{Few-random shot (5)}
& AUC & $0.810 \pm 0.002$ & $0.785 \pm 0.005$ & $0.801 \pm 0.003$ & $0.786 \pm 0.017$ & $0.837 \pm 0.005$ & $0.775 \pm 0.002$ & $\mathbf{0.850 \pm 0.006}$ & $0.803 \pm 0.007$ & $0.810 \pm 0.019$ & $0.776 \pm 0.005$ \\
& Accuracy & $0.523 \pm 0.003$ & $0.435 \pm 0.005$ & $0.513 \pm 0.016$ & $0.443 \pm 0.014$ & $0.503 \pm 0.013$ & $0.435 \pm 0.007$ & $\mathbf{0.570 \pm 0.016}$ & $0.521 \pm 0.018$ & $\mathbf{0.614 \pm 0.035}$ & $0.500 \pm 0.018$ \\
& AOD & $0.818 \pm 0.003$ & $0.782 \pm 0.003$ & $0.829 \pm 0.009$ & $0.790 \pm 0.008$ & $0.823 \pm 0.005$ & $0.770 \pm 0.002$ & $\mathbf{0.837 \pm 0.005}$ & $0.819 \pm 0.006$ & $\mathbf{0.852 \pm 0.011}$ & $0.765 \pm 0.012$ \\
\addlinespace

\multirow{3}{*}{\shortstack[l]{Documentation-based\\Few-shot (5)}}
& AUC & $0.808 \pm 0.004$ & $\mathbf{0.813 \pm 0.005}$ & - & - & - & - & - & - & - & - \\
& Accuracy & $0.504 \pm 0.004$ & $0.483 \pm 0.007$ & - & - & - & - & - & - & - & - \\
& AOD & $0.806 \pm 0.003$ & $0.800 \pm 0.001$ & - & - & - & - & - & - & - & - \\

\bottomrule
\end{tabular}}%
\end{table*}

\begin{table*}[t]
\centering
\caption{Summary of the performance of the functionality-only retrieval compared to other approaches (metrics are obtained from 5 bootstrap runs and for CodeLlama-7B). Documentation-based retrieval is only applicable to the two projects that define components in their documentation; a dash indicates the configuration is not applicable for that project.}
\label{tab:full_functionality}
\small
\resizebox{\textwidth}{!}{\begin{tabular}{l l c c c c c c c c c c}
\toprule
\textbf{Model} & \textbf{Metric} & \textbf{Hadoop} & \textbf{HBase} & \textbf{Elasticsearch} & \textbf{Cassandra} & \textbf{Flink} & \textbf{Kafka} & \textbf{JMeter} & \textbf{ZooKeeper} & \textbf{Karaf} & \textbf{Wicket} \\
\midrule

\multirow{3}{*}{\textbf{Functionality-based (5)}}
& AUC & $\mathbf{0.865 \pm 0.002}$ & $\mathbf{0.863 \pm 0.003}$ & $\mathbf{0.924 \pm 0.002}$ & $\mathbf{0.863 \pm 0.006}$ & $\mathbf{0.860 \pm 0.004}$ & $\mathbf{0.831 \pm 0.008}$ & $\mathbf{0.859 \pm 0.008}$ & $\mathbf{0.834 \pm 0.013}$ & $\mathbf{0.856 \pm 0.060}$ & $\mathbf{0.798 \pm 0.029}$ \\
& Accuracy & $\mathbf{0.587 \pm 0.002}$ & $\mathbf{0.568 \pm 0.007}$ & $\mathbf{0.638 \pm 0.008}$ & $\mathbf{0.652 \pm 0.020}$ & $\mathbf{0.556 \pm 0.008}$ & $\mathbf{0.541 \pm 0.019}$ & $\mathbf{0.594 \pm 0.012}$ & $\mathbf{0.564 \pm 0.019}$ & $\mathbf{0.629 \pm 0.038}$ & $\mathbf{0.523 \pm 0.039}$ \\
& AOD & $\mathbf{0.837 \pm 0.003}$ & $\mathbf{0.838 \pm 0.004}$ & $\mathbf{0.863 \pm 0.001}$ & $\mathbf{0.868 \pm 0.005}$ & $\mathbf{0.839 \pm 0.004}$ & $\mathbf{0.818 \pm 0.009}$ & $\mathbf{0.840 \pm 0.007}$ & $\mathbf{0.840 \pm 0.010}$ & $\mathbf{0.858 \pm 0.016}$ & $\mathbf{0.788 \pm 0.023}$ \\
\addlinespace

\multirow{3}{*}{Few-random shot (5)}
& AUC & $0.810 \pm 0.002$ & $0.785 \pm 0.005$ & $0.801 \pm 0.003$ & $0.786 \pm 0.017$ & $0.837 \pm 0.005$ & $0.775 \pm 0.002$ & $0.850 \pm 0.006$ & $0.803 \pm 0.007$ & $0.810 \pm 0.019$ & $0.776 \pm 0.005$ \\
& Accuracy & $0.523 \pm 0.003$ & $0.435 \pm 0.005$ & $0.513 \pm 0.016$ & $0.443 \pm 0.014$ & $0.503 \pm 0.013$ & $0.435 \pm 0.007$ & $0.570 \pm 0.016$ & $0.521 \pm 0.018$ & $0.614 \pm 0.035$ & $0.500 \pm 0.018$ \\
& AOD & $0.818 \pm 0.003$ & $0.782 \pm 0.003$ & $0.829 \pm 0.009$ & $0.790 \pm 0.008$ & $0.823 \pm 0.005$ & $0.770 \pm 0.002$ & $0.837 \pm 0.005$ & $0.819 \pm 0.006$ & $0.852 \pm 0.011$ & $0.765 \pm 0.012$ \\
\addlinespace

\multirow{3}{*}{\shortstack[l]{Documentation-based\\Few-shot (5)}}
& AUC & $0.808 \pm 0.004$ & $0.813 \pm 0.005$ & - & - & - & - & - & - & - & - \\
& Accuracy & $0.504 \pm 0.004$ & $0.483 \pm 0.007$ & - & - & - & - & - & - & - & - \\
& AOD & $0.806 \pm 0.003$ & $0.800 \pm 0.001$ & - & - & - & - & - & - & - & - \\

\bottomrule
\end{tabular}}%
\end{table*}

\begin{table*}[t]
\centering
\caption{Impact of the different retrieval strategies on the CodeLlama-34B performance. The purpose of the 34B evaluation is to establish how our results scale with model size, and that question is answered by the multiplex row, which is the configuration we recommend. The ablation of the individual retrieval signals is a question about the retrieval mechanism rather than about model capacity, and we therefore answer it on the CodeLlama-7B model (Tables~\ref{tab:full_ownership} and~\ref{tab:full_functionality}). A dash indicates the configuration was not evaluated for that project.}
\label{tab:appendix_34b_ablation}
\small
\resizebox{\textwidth}{!}{%
\begin{tabular}{l l c c c c c c c c c c}
\toprule
\textbf{Retrieval} & \textbf{Metric} & \textbf{Hadoop} & \textbf{HBase} & \textbf{Elastic.} & \textbf{Cassandra} & \textbf{Flink} & \textbf{Kafka} & \textbf{JMeter} & \textbf{ZooKeeper} & \textbf{Karaf} & \textbf{Wicket} \\
\midrule
\multirow{3}{*}{Functionality}
& AUC       & 0.875 $\pm$ 0.001 & 0.871 $\pm$ 0.005 & 0.889 $\pm$ 0.037 & 0.859 $\pm$ 0.009 & - & - & - & - & - & - \\
& Accuracy  & 0.616 $\pm$ 0.008 & 0.580 $\pm$ 0.012 & 0.644 $\pm$ 0.004 & 0.569 $\pm$ 0.026 & - & - & - & - & - & - \\
& AOD       & 0.850 $\pm$ 0.004 & 0.838 $\pm$ 0.007 & 0.865 $\pm$ 0.001 & 0.844 $\pm$ 0.011 & - & - & - & - & - & - \\
\addlinespace
\multirow{3}{*}{Ownership}
& AUC       & 0.824 $\pm$ 0.002 & 0.812 $\pm$ 0.006 & 0.905 $\pm$ 0.001 & 0.838 $\pm$ 0.012 & - & - & - & - & - & - \\
& Accuracy  & 0.544 $\pm$ 0.004 & 0.508 $\pm$ 0.018 & 0.582 $\pm$ 0.005 & 0.504 $\pm$ 0.020 & - & - & - & - & - & - \\
& AOD       & 0.822 $\pm$ 0.001 & 0.809 $\pm$ 0.007 & 0.843 $\pm$ 0.003 & 0.821 $\pm$ 0.010 & - & - & - & - & - & - \\
\addlinespace
\multirow{3}{*}{\textbf{Multiplex}}
& AUC       & \textbf{0.932 $\pm$ 0.006} & \textbf{0.926 $\pm$ 0.003} & \textbf{0.969 $\pm$ 0.001} & \textbf{0.914 $\pm$ 0.002} & \textbf{0.887 $\pm$ 0.005} & \textbf{0.860 $\pm$ 0.004} & \textbf{0.877 $\pm$ 0.013} & \textbf{0.879 $\pm$ 0.021} & \textbf{0.879 $\pm$ 0.033} & \textbf{0.836 $\pm$ 0.013} \\
& Accuracy  & \textbf{0.727 $\pm$ 0.008} & \textbf{0.694 $\pm$ 0.008} & \textbf{0.775 $\pm$ 0.005} & \textbf{0.693 $\pm$ 0.003} & \textbf{0.589 $\pm$ 0.012} & \textbf{0.585 $\pm$ 0.009} & \textbf{0.652 $\pm$ 0.022} & \textbf{0.593 $\pm$ 0.034} & \textbf{0.699 $\pm$ 0.012} & \textbf{0.552 $\pm$ 0.049} \\
& AOD       & \textbf{0.894 $\pm$ 0.005} & \textbf{0.880 $\pm$ 0.003} & \textbf{0.913 $\pm$ 0.001} & \textbf{0.884 $\pm$ 0.003} & \textbf{0.850 $\pm$ 0.005} & \textbf{0.826 $\pm$ 0.006} & \textbf{0.857 $\pm$ 0.010} & \textbf{0.840 $\pm$ 0.018} & \textbf{0.873 $\pm$ 0.007} & \textbf{0.811 $\pm$ 0.025} \\
\bottomrule
\end{tabular}%
}
\end{table*}

\end{document}